\documentclass{article}
\usepackage{graphicx}
\usepackage{amsfonts}
\usepackage{floatrow}
\usepackage{authblk}
\usepackage{cite}
\usepackage{bm}
\usepackage{amsmath}
\usepackage{amssymb}

\newcommand{\mat}[2]{\left[ \begin{array}{*{#1}{c}}#2\end{array}\right]}

\title{An algebraic approach to FQHE variational wave functions}
\author{S. B. Mulay}
\affil{\small Department of Mathematics, University of Tennessee, Knoxville, TN 37996, USA}
\author{J. J. Quinn}
\affil{\small Department of Physics, University of Tennessee, Knoxville, TN 37996, USA}
\author{M. A. Shattuck}
\affil{\small Department of Mathematics, University of Tennessee, Knoxville, TN 37996, USA}

\begin{document}
\maketitle
\noindent \emph{Abstract}\\

Consider a system of $N$ electrons projected onto the lowest Landau level (LLL) with filling factor of the form $\frac{n}{2pn\pm1}<\frac{1}{2}$ and $N$ a multiple of $n$.  We show that there always exists a two-dimensional symmetric correlation factor (arising as a nonzero symmetrization) for such systems and hence one can always write a variational wave function. This extends an earlier observation of Laughlin for an incompressible quantum liquid (IQL) state with filling factor equal to the reciprocal of an odd integer $\geq$ 3.  To do so, we construct a family of $d$-regular multi-graphs on $N$ vertices for any $N$ whose graph-monomials have nonzero linear symmetrization and obtain, as special cases, the aforementioned nonzero correlations for the IQL state.  The nonzero linear symmetrization that is obtained is in fact an example of what is called a binary invariant of type $(N,d)$.  Thus, in addition to supplying new variational wave functions for systems of interacting Fermions, our construction is of potential interest from both the graph and invariant theoretic viewpoints.\\



\section{Introduction}

A trial wave function $\Psi(z_1,\ldots,z_N)$ of a fractional quantum Hall effect (FQHE) system of $N$ electrons can always be expressed as the product of an
antisymmetric Fermion factor $F = \prod_{1 \leq i < j \leq N}(z_i-z_j)$, and a symmetric correlation factor $G=G(z_1,\ldots,z_N)$ that takes into account Coulomb interactions.   In this paper, we will be interested in certain mathematical aspects of the latter.  Here, and in the following, let $z_{i}$ denote the complex coordinate of the $i^{\rm{th}}$ electron.  $G$ in general is assumed to be an analytic function of the $z_i$ and here will be taken to be a homogeneous polynomial in these variables.  We refer to the difference $z_i-z_j$ as a correlation factor (cf), even when it arises from the Pauli principle. For non-interacting Fermion systems, one may take $G = 1$.  It is convenient to represent a configuration of electrons diagrammatically as a multi-graph on $N$ vertices with edges representing cf factors.

We discuss some previous examples of correlations $G$.  In an incompressible quantum liquid (IQL) state with filling factor $\nu = {^1/_3}$ (see \cite{Laughlin83}), two cf lines
connect each pair of Fermions.  For each labeling of the vertices of the corresponding multi-graph, one takes the product of all the cf factors, and then computes the sum of the products corresponding to all possible labelings to obtain the correlation $G$. For the Moore-Read state \cite{Moore91} of the half filled first excited Landau level (LL1) with $\nu = 2 + {^1/_2}$,
the $N$ electrons for LL1 where $N$ is even are partitioned into two subsets $A$ and $B$, each of size $m=N/2$,
with two cf's joining each pair of electrons in $A$ and also each pair in $B$, as noted in \cite{Quinn14}. To compute the correlation $G$ in this case, one takes the product of all cf factors in the diagram corresponding to a given partition $(A,B)$, and then sums these products over the possible partitions to obtain $G$.  The Jain ${^2/_5}$ IQL state can be represented in a similar fashion as the Moore-Read state above except that an extra $m(m-1)$ intersubset cf's are required between the particles in the subsets $A$ and $B$ so as to ensure that the total angular momentum $L$ is zero (see \cite{Quinn14}), as illustrated below in the case $N=4$.  Jain \cite{Jain89, Jain90} introduced a more general composite Fermion (CF) picture that correctly predicts the IQL states at filling factors $<$ ${^1/_2}$, which correspond to integrally filled CF Landau levels.

\begin{figure}[h!]
\centering
\includegraphics[scale=0.4]{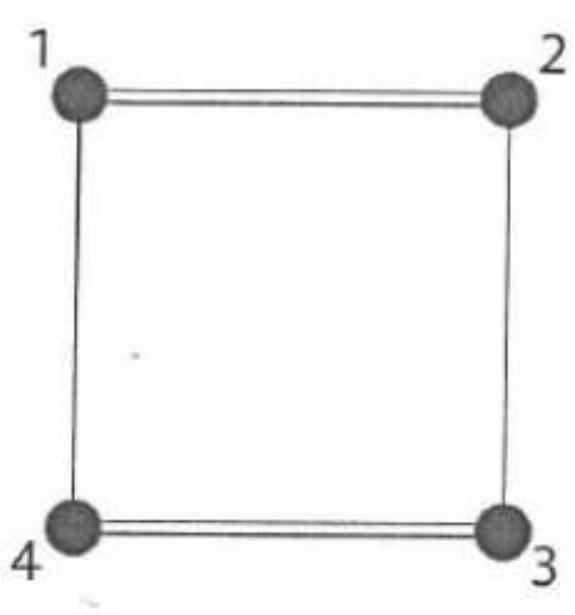}
\caption{\small Jain 2/5 IQL state with $N=4$.}
\end{figure}

The value of $(2l, N)$ defines the function space consisting of the $2l + 1 $ states into which one must insert $N$ Fermions, where $l$ is the single particle angular momentum (for noninteracting single particle states); see, e.g., \cite[Chapter 16]{QY} or \cite{BQ}.  Here, $l$ may be either an electron or a quasielectron angular momentum.   The filling factor is defined as $\nu=\frac{N}{2l+1}$, where $N$ is assumed large (formally, $\nu$ is the limit as $N \rightarrow \infty$, though it is often used for systems with large numbers of particles).  In general, for states at filling factor $\nu = p/q < {^1/_2}$ and $N$ arbitrary, the value of the single
particle angular momentum $l$ satisfies the relation $2l = \nu^{-1}N - c_\nu$, where $c_\nu = q + 1 - p$ is the finite size shift ($c_\nu$ serves as a correction term for small $N$ since one might not wish to use the limiting filling factor value $\nu$ in this case).

The correlation $G$ satisfies a number of conditions. For example, the highest power of $z_i$ in any term of $G$ cannot exceed $2l + 1 - N$. In addition, the value $L$ of the total angular momentum of the correlated state must satisfy
the equation $L = \left(N/2\right)\,\left(2l+1-N\right)- \kappa_G$, where $\kappa _G$ is the degree of the homogeneous polynomial
$G$. Knowing the value of $L$ for IQL states and for states containing a few quasielectrons (or a few quasiholes) from Jain's
mean field CF picture allows one to determine $\kappa _G$. In IQL states with filling factors $\nu<{^1/_2}$, which we will focus on here, one has $L=0$ with the highest power of $z_i$ in $G$ generally attained for all $i$, which we will denote by $d$.  By definition, the total angular momentum gives the degree of the homogeneity of the polynomial $G$ in LLL.  Here, note that we have $L=0$ without $G$ being a constant since we are dealing with composite Fermion wave functions where one starts with functions in higher Landau levels (such functions can have $L=0$ without being constant) and projects them onto LLL.

Let $\nu=\frac{n}{2pn\pm1}$, where $n,p\geq 1$ are integers.  The degree $\kappa_G$ then satisfies $\kappa_G=\frac{Nd}{2}$, where
$$d=2\ell+1-N=\nu^{-1}N-c_\nu+1-N=((2 p - 1) n \pm 1)((N / n) - 1).$$

Laughlin~\cite{Laughlin83} realized that if the interacting electrons could avoid the most strongly repulsive pair states, an incompressible quantum liquid state could result.
He suggested a trial wave function for a filling factor $\nu$ equal to the reciprocal of an odd integer $n$ in which the correlation, denoted by $G_L$, was given by $\prod_{1\leq i<j\leq N}{(z_i-z_j)^{n-1}}$.
One can represent the configuration for $G_L$ diagrammatically by distributing $N$ dots, representing $N$ electrons along the circumference of a circle, and drawing $\frac{n-1}{2}$ double lines, each representing two correlation factors connecting each pair.  Note that $G_L$ is given by an integral power of the discriminant of the $z_i$ and hence is nonzero.

In general, in order for a given configuration to exist,  it is necessary that the symmetric correlation factor $G$ work out to be nonzero, as it does in the Laughlin case.  Note that if a configuration contains pairs of vertices connected by an odd number of cf lines, this is often not obvious.  Here, we consider a general class of configurations of $N$ Fermions (Laughlin's arrangement being a special case of which) containing members of all IQL states with $v<{^1/_2}$ of the stated form above and study the correlations $G$ corresponding to this class.  Members belonging to a certain subclass of these configurations and exhibiting two kinds of pair-interaction potencies $a$ and $b$ are described as \emph{balanced}, with the \emph{minimal} configurations being those in which $\max\{a,b\}$ is as small as possible (see final section for further details).  The following result is a consequence of Theorems 2-4 below and their application to the IQL state.\\

\noindent \underline {\bf Theorem 1}: Let $N$, $n$ and $p$ be positive integers where $N$ is a multiple of $n$.  Then for all $N$, $n$ and $p$ such that $\nu=n/(2pn\pm1)<{^1/_2}$, there exist configurations of $N$ Fermions in the IQL state with filling factor $\nu$ that have a nonzero symmetric correlation factor $G$. Furthermore, it is possible to find for all such $\nu$ configurations that are minimal in the sense defined in the final section below.  \\

\noindent Note that the $n=1$ case of Theorem 1 corresponds to the Laughlin configuration having nonzero $G_L$.

The organization of this paper is as follows.  In section II, we discuss the algebra of variational wave functions and formulate the problem in terms of multi-graphs.  It will be seen that the filling factor and angular momentum requirements for a system of electrons imposes certain conditions on the associated multi-graph and thus on $G$.  In the third section, we introduce notation and recall some terminology.  We present in section IV our main results featuring the construction of certain kinds of invariants.  In the final section, we discuss applications of our results to the IQL state and show how Theorem 1 above follows as a consequence.  Further, we demonstrate that our construction yields existent configurations in which the maximal occurring pair-interaction potency is at its lowest possible value (among the possible configurations exhibiting two types of potency and having the given filling factor).

\section{Preliminaries}

Recall that
a correlation diagram for $N$ Fermions graphically exhibits the potencies
of their mutual interactions and so, in purely mathematical terms, it is
a (undirected, loopless) {\it multi-graph} on $N$ vertices. Here, the adjective
{\it multi-} signifies the possibility that a vertex-pair may be connected by more
than one edge. Henceforth, we regard {\it correlation diagram} and {\it multi-graph}
as equivalent terms. Given a multi-graph on $N$ vertices, a choice of a labeling
of its vertices by the numbers $1, 2, \dots, N$ gives rise to a product of the terms
$(z_{i} - z_{j})^{p_{ij}}$, where $z_{i}$ is an indeterminate for $1 \leq i \leq N$ and
for $1 \leq i < j \leq N$, the nonnegative integer $p_{ij}$ is the number of edges between
the vertices labeled $i$ and $j$ in the multi-graph. In the classical theory of invariants,
a product of this type is known as a {\it graph-monomial} (see, e.g., \cite{AS}). Note that since our $N$ Fermions are
indistinguishable, we must consider each of the possible choices of vertex-labelings, for the
correlation diagram under consideration, on an equal footing. Commonly, two multi-graphs
on $N$ vertices are called {\it isomorphic} if one is obtained from the other by a
relabeling of its vertices (see Figure 2 below for an example of isomorphic multi-graphs).

\begin{figure}[t]
\centering
\begin{tabular}{cc}
\includegraphics[scale=0.3]{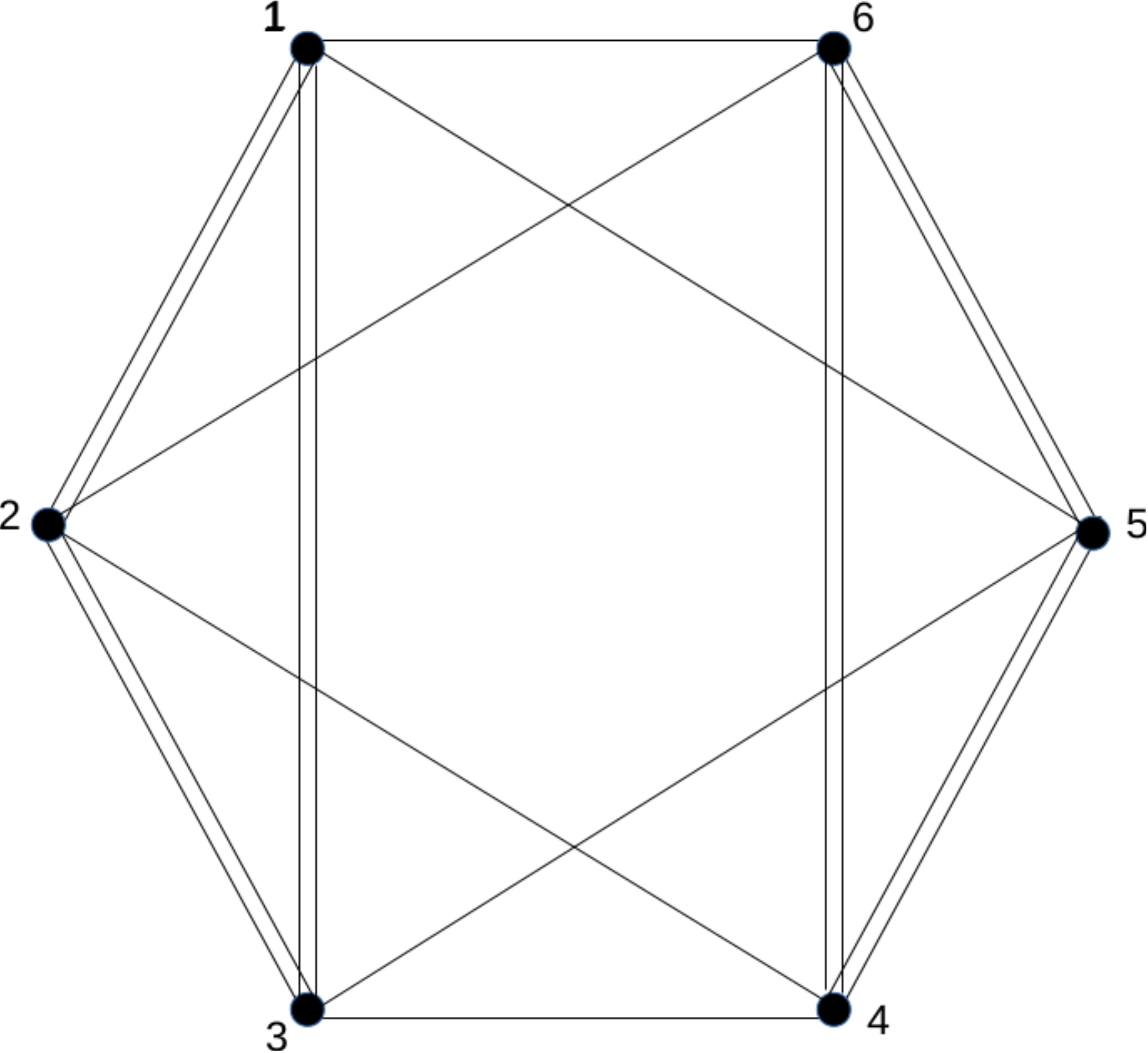} & \includegraphics[scale=0.3]{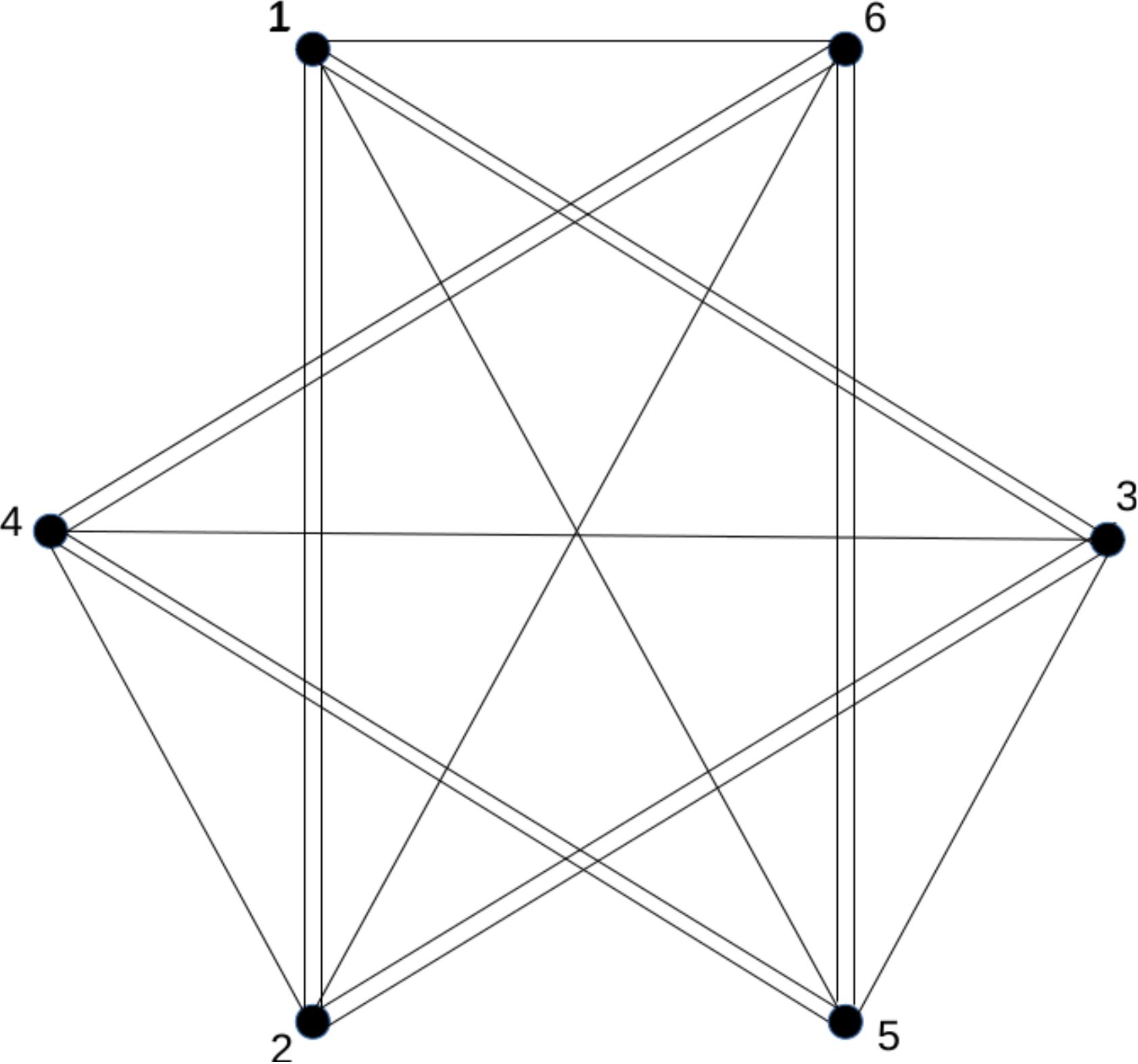}
\end{tabular}
\caption{\small Isomorphic multi-graphs on $6$ vertices.}
\end{figure}

The isomorphism class of a correlation diagram is to be thought of as a
{\it configuration} of interacting Fermions; nonisomorphic correlation diagrams
correspond to distinct configurations. The sum (if
preferred, it can also be defined as the average) of the graph-monomials associated with
a configuration of $N$ interacting Fermions will be referred to here as the \emph{correlation function} (of the configuration). In other words, if we pick one correlation diagram for the
configuration and call its associated graph-monomial $f(z_{1}, \dots, z_{N})$, then
the correlation function of the configuration is the {\it symmetrization} of $f$,
{\it i.e.}, $\sum f(z_{\sigma (1)}, \dots , z_{\sigma (N)})$, where the sum ranges
over all permutations $\sigma$ of $\{1, 2, \dots, N \}$. Clearly, such a correlation
function is a homogeneous polynomial symmetric in $z_{1}, z_{2}, \dots , z_{N}$. If
this correlation function is identically zero, then we deem the configuration as
{\em nonexistent}. If correlation functions of two configurations are the same
up to a nonzero numerical (rational) factor, then the configurations are regarded as
{\it equivalent}.

It is worth noting that on account of the inherent symmetries of a given multi-graph, it
can very well be the case that certain distinct labelings of vertices yield the same
graph-monomial. From a computational standpoint, the correlation function of a configuration
is easier to deal with when its corresponding set of graph-monomials is small and hence
multi-graphs with many intrinsic symmetries are perhaps more desirable. In the extremal example of a
multi-graph in which the number of edges between any two vertices is the same integer $e$
({\it i.e.}, $p_{ij} = e$ for $1 \leq i < j \leq N$), there are at most two distinct
graph-monomials for the configuration (differing only by a factor of $\pm 1$). Recall that
such is precisely the case if we consider the Laughlin configuration for the IQL state with filling
factor $\nu = 1 / (2 p + 1)$ (forcing $p_{ij} = e = 2 p$). In general, a simple exercise shows that
the graph-monomial of a multi-graph on $N$ vertices is a symmetric polynomial in the variables
$z_{1}, z_{2}, \dots , z_{N}$ if and only if there is an integer $p$ such that $p_{ij} = 2 p$ for
all $1 \leq i < j \leq N$.

For a system of $N$ interacting Fermions, their individual angular momenta, together with
the filling factor $\nu$, dictate an upper bound $d$ on the degree of a vertex ({\it i.e.}, the
number of edges emanating from a vertex) in any corresponding correlation diagram, whereas the
total angular momentum $L$ of the system demands that the corresponding correlation function
be a homogeneous polynomial of (total) degree $(N d / 2) - L$. Usually, there are several possible
configurations that meet these dictated requirements; their number increases rather steeply with
increasing values of $N$. To determine which of these configurations actually exist, it is
essential to ascertain the nonzero-ness of their corresponding correlation functions. This is a
nontrivial task when the associated correlation diagram has vertex-pairs connected by an odd number
of edges. Even more challenging is the problem of determining, in some concrete manner, the set of
equivalence classes of these configurations.

The simplest, but comparatively rare, example that can be
worked out by hand is given by a system of $4$ Fermions in an IQL state with filling factor ${^2/_5}$;
in this case, each vertex of a correlation diagram must have degree $3$ and
then there is only one existent configuration.  In general, if $L = 0$, then it turns out that each vertex
in a related correlation diagram must have the same maximum allowed degree $d$. A (undirected, loopless)
multi-graph each of whose vertices has the same degree $d$ is said to be {\it $d$-regular}. The problem of
counting the number of distinct configurations of $N$ Fermions with $L = 0$ and a given filling factor $\nu$
translates to counting the number of isomorphism classes of $d$-regular loopless multi-graphs on $N$ vertices.
We wish to point out that this counting problem appears to be largely open and is a subject of ongoing research
(see~\cite{GM}).

For a system of $N$ Fermions in an IQL state with filling factor $\nu = n / (2 p n \pm 1) \; < {^1/_2}$,
we have $L = 0$ and $d = ((2 p - 1) n \pm 1)((N / n) - 1)$. In particular, if $\nu = {^1/_3}$ and hence $d = 2 (N  - 1)$,
thanks to {\bf gtools}, {\bf MAPLE}, and {\bf SAGE}, we can present at least a small sample of relevant counts in Table 1 above.  Two configurations with $N=6$ and $\nu={^2/_5}$ (and hence $d=6$) are given below.
\begin{table}
\begin{center}
\begin{tabular}{|c|c|c|c|c|c|c|}
\hline no. of Fermions & $2$ & $3$ & $4$ & $5$ & $6$ & $7$ \\ \hline
no. of apparent configurations & $1$ & $1$ & $7$ & $37$ & $2274$ & $864863$ \\ \hline
no. of existent configurations & $1$ & $1$ & $7$ & $33$ & $1137$ & $844578$ \\ \hline
\end{tabular}
\caption{\small Apparent and existent configurations, IQL state, $\nu = 1 / 3$.}
\end{center}
\end{table}

\begin{figure}[h]
\centering
\begin{tabular}{cc}
\includegraphics[scale=0.4]{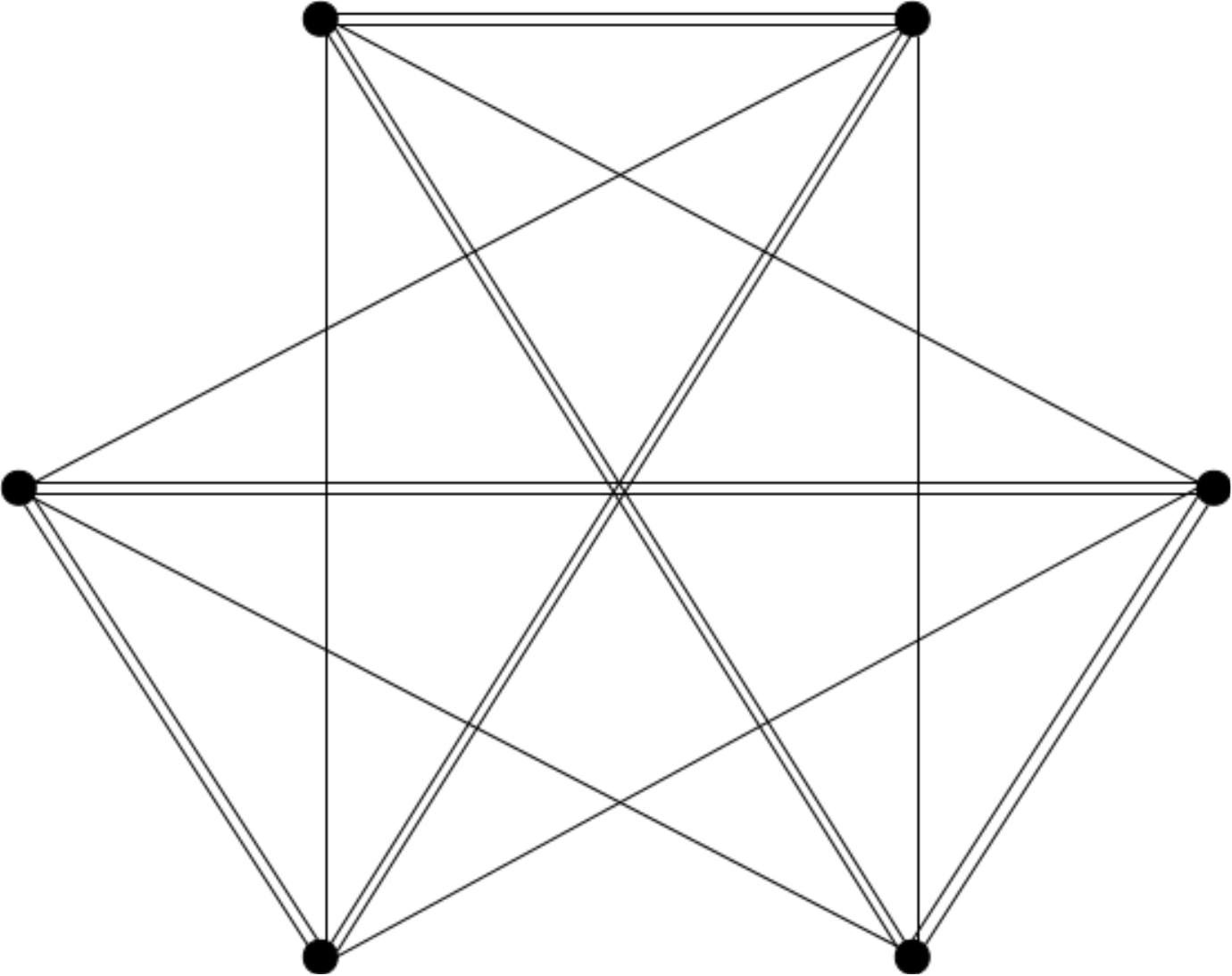} & \includegraphics[scale=0.4]{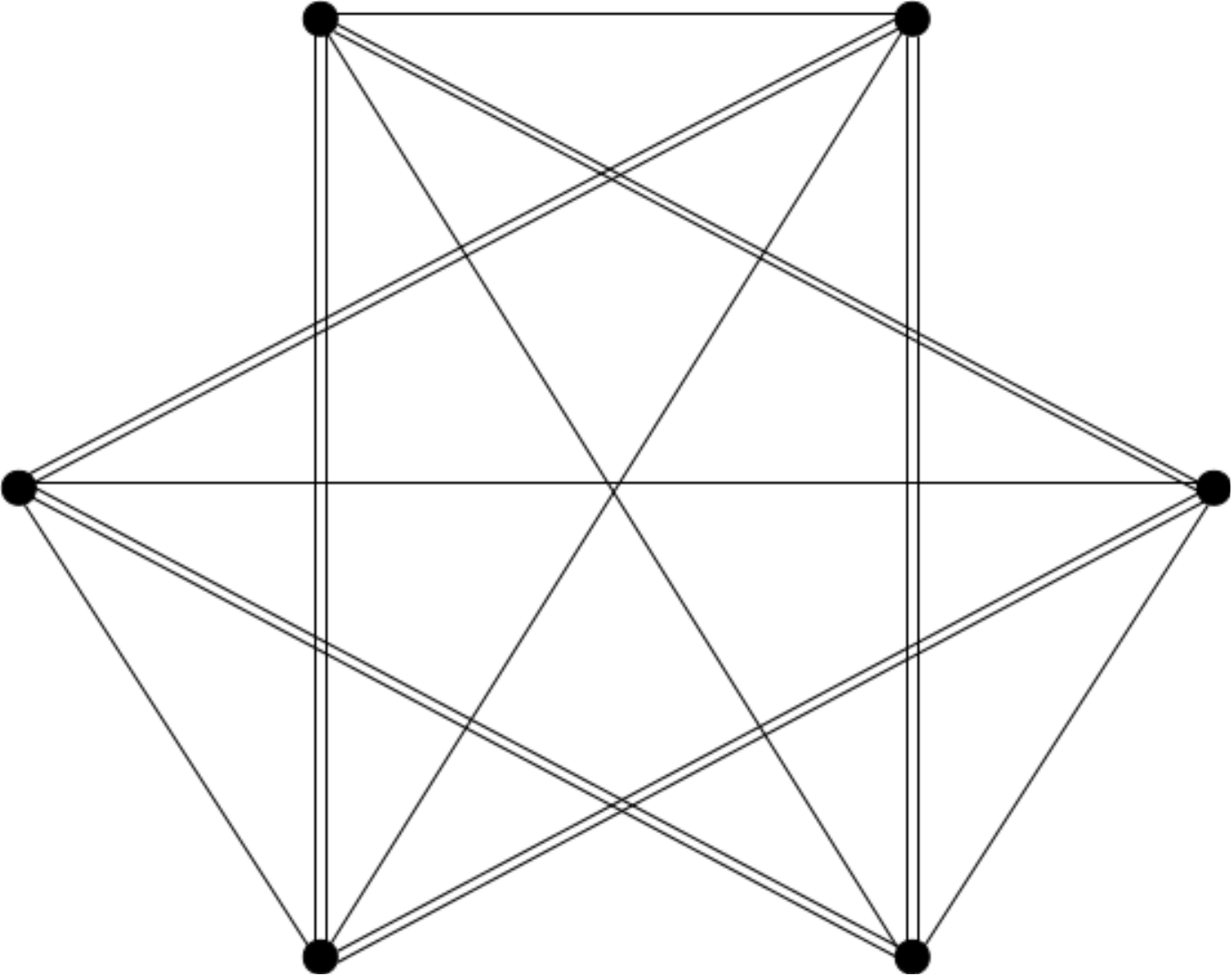}
\end{tabular}
\caption{\small Distinct configurations for $N=6$ Fermions in the IQL state with $\nu = 2/5$.}
\end{figure}

To illustrate one of our subsequent results, let $N=mn$ with  $m=n=3$ and $p=1$ and hence $d=((2p-1)n +1)(m - 1)=8$.  Note that this corresponds to an IQL state with $N=9$ and $\nu=3/7$, as shown in Figure 4.  In the notation of the fourth section, this graph has adjacency matrix $M(3,3,1,1)$ whose symmetrized graph monomial is proven to be nonzero (see Corollary to Theorem 3 below), which implies that the configuration is indeed existent.

\begin{figure}[h]
\centering
\includegraphics[scale=0.4]{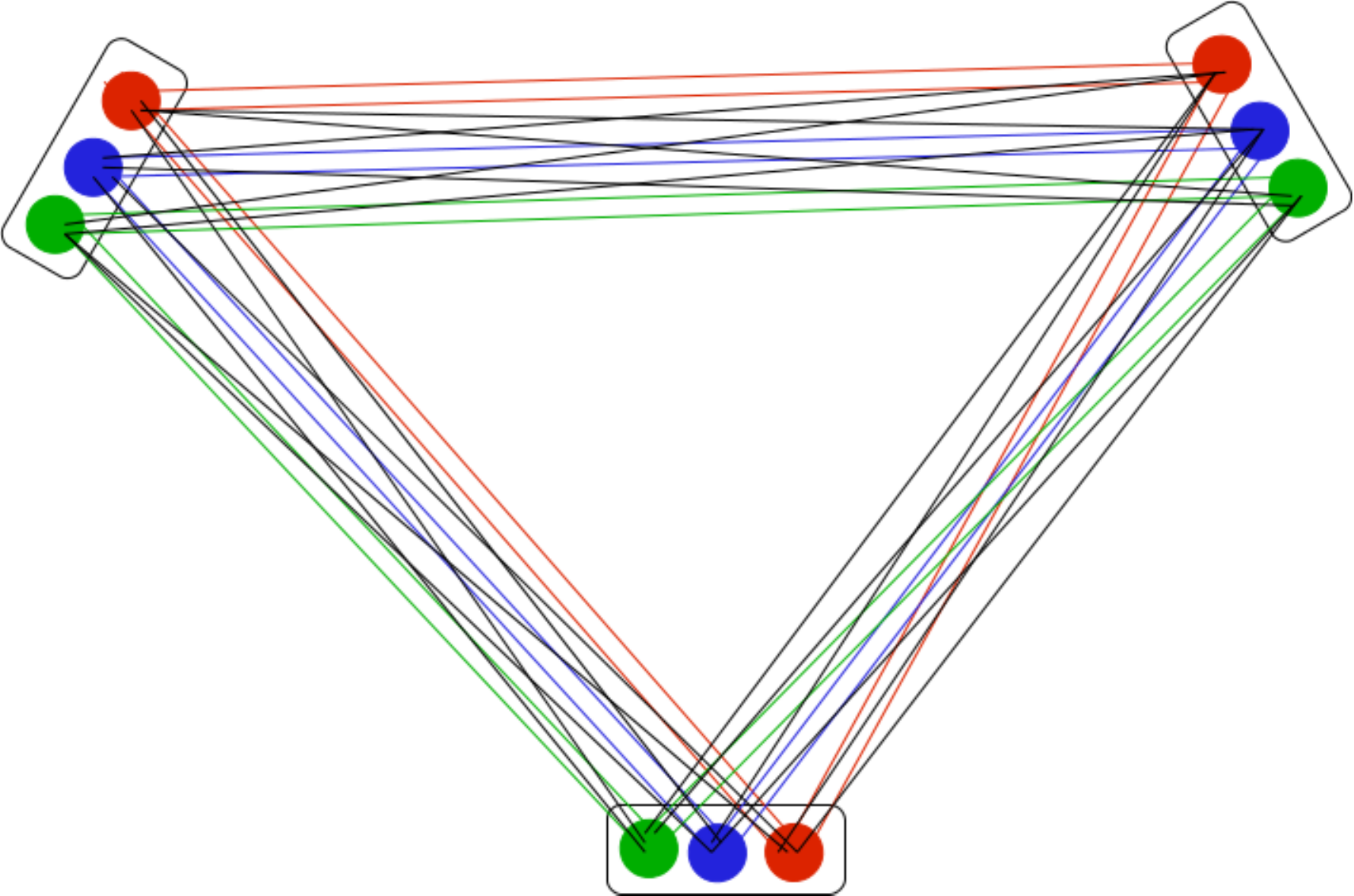}
\caption{\small An existent IQL state with $N=9$ and $\nu=3/7$.}
\end{figure}

It is well-known that the symmetrized graph-monomial of an undirected loopless multi-graph is
also called a {\it relative semi-invariant} of a (generic) binary form of degree $N$. If the multi-graph is $d$-regular, then the associated symmetrized graph-monomial is a {\it
relative invariant} of the degree $N$ binary form. What is of key interest in our context
is the fact that the symmetrized graph-monomials of the $d$-regular multi-graphs on $N$ vertices
constitute a generating set for the vector space (over a field of characteristic $0$) formed by
the relative invariants of degree $d$ and weight $N d/ 2$. So, for a system of $N$
interacting Fermions with filling factor $\nu$ and $L = 0$, the set of all correlation functions of
the corresponding configurations generates the vector space of the relative invariants of weight
$N d / 2$, where $d$ is the aforementioned bound dictated by the parameters of the system. For such
a system, it follows that we can hope to find at least one nonzero correlation function if and only
if this vector space is nonzero, and the configurations associated to our system form a single
equivalence class if and only if the vector space is one-dimensional. Since a generating function
for dimensions of vector spaces of relative invariants has been
known for over a century (see \cite[Articles 183--187]{E}) and a complete list of pairs $(N, N d / 2)$ for which the corresponding space
of invariants is nonzero has been recently determined (see \cite[Proposition 4.2]{BDP} and \cite{D}), the observations above are
indeed useful. Ever since Cayley founded the theory of invariants, explicit construction of (semi-) invariants has been of extensive interest. Though our motivation for the explicit constructions of invariants formulated in the subsequent theorems lies in building correlation functions for systems of interacting Fermions, these theorems have more to offer from a purely invariant theoretic point of view.  For a deeper, more comprehensive treatment of the theory of invariants of binary forms, the interested reader may wish to consult either the classic~\cite{GY} or the more contemporary exposition~\cite{KR}.

\section{Notation and terminology}

Although multi-graphs can be visually pleasing, it is undoubtedly simpler to deal with their adjacency matrices
in attempting to prove precise results. Thus the reader will find our definitions and theorems formulated in the language of matrices.

In what follows, ${\Bbb Z}$ denotes
the set of ordinary integers, ${\Bbb N}$ denotes the set of nonnegative
integers and ${\Bbb Q}$ denotes the set of rational numbers. For a function $f$
defined on a set $S$, by $f(S)$, we mean the set $\{ f(a) \mid a \in S \}$. For
the cardinality of a set $S$, we use the notation $|S|$. In this paper, we are mainly interested in polynomials
and rational functions having coefficients in an integral domain of characteristic zero; nevertheless,
some of the definitions, examples and proofs may remain valid in the positive characteristic case.
Since {\it degree} and {\it order} of a rational function play an important role in our
proofs, it is helpful to briefly recall their definitions and properties. Consider a rational
function $f$ in a set of indeterminates $z$ such that $f = P / Q$ for some nonzero polynomials $P$ and $Q$
in $z$ having coefficients in an integral domain $k$. Then the degree of $f$ is defined to be the difference
between the (usual) degrees of $P$ and $Q$. By convention, $0$ has degree $ - \infty$. Let $g$ be also
a rational function in $z$ with coefficients in $k$. Recall that the degree of $f g$ is the sum of the
degrees of $f$ and $g$, whereas the degree of $f + g$ is bounded above by the maximum of the degrees of $f$ and
$g$. Moreover, the degree of $f + g$ is the maximum of the degrees of $f$ and $g$ whenever $f$ and $g$ have
unequal degrees. Now suppose $k$ is a unique factorization domain and $J$ is a nonzero principal prime ideal of
the polynomial ring $k [z]$. Then, the {\it $J$-order} of a nonzero polynomial $h \in k[z]$ is defined to be the largest
nonnegative integer $m$ such that $h$ is in $J^{m}$. Subsequently, the $J$-order of $f$ is defined to be the
difference between the $J$-orders of $P$ and $Q$. By convention, the $J$-order of $0$ is $\infty$. If $u \in k[z]$
is a generator of $J$, then the term {\it $u$-order} is regarded to be synonymous with the term $J$-order. Recall
that the $J$-order of $f g$ is the sum of their respective $J$-orders whereas the $J$-order of $f + g$ is bounded
below by the minimum of the $J$-orders of $f$ and $g$. Moreover, the $J$-order of $f + g$ equals the minimum of the
$J$-orders of $f$ and $g$ whenever $f$ and $g$ have unequal $J$-orders. For various other notions from basic abstract
algebra that are tacitly used in the rest of this article, the reader is referred to \cite{SZ}. \\  \\

\section{Construction of invariants}

\noindent \underline {\bf Definitions}: Let $N\geq2$ be an integer.
As before, let $k$ be a field containing ${\Bbb Q}$,
let $z_{1}, \dots, z_{N}$ be indeterminates and let $z$ stand for
$(z_{1}, \dots , z_{N})$.
\begin{enumerate}
\item By $S_{N}$, we denote the group of all permutations of $\{1 , \ldots, N \}$.
For any ring $k$, let
\[ Symm_{N} \, : k [z_{1}, \dots , z_{N}] \rightarrow k [z_{1}, \dots , z_{N}] \]
be the {\it Symmetrization operator} given by
\[ Symm_{N} (f(z_{1}, \ldots , z_{N})) :=
\sum_{\sigma \in S_{N}} f(z_{\sigma (1)}, \ldots , z_{\sigma (N)}) .\]
$f$ is {\it symmetric} provided
$f(z_{\sigma (1)}, \ldots , z_{\sigma (N)}) = f(z_{1}, \ldots , z_{N})$ for all
$\sigma \in S_{N}$.
\item Given an $N \times N$ matrix $A := [a_{ij}]$ with integer entries, let
$r_{i}$ denote the sum of the entries in the $i$-th row of $A$ for $1 \leq i \leq N$
and define
\[ \rho (A)\, : = \, (r_{1}, \ldots , r_{N}) \, . \]
\item Given an $N \times N$ matrix $A := [a_{ij}]$, where each $a_{ij}$ is a nonnegative
integer, letting $z$ stand for the vector $(z_{1}, \ldots , z_{N})$, define
\[ \delta (z,\, A) \, := \, \prod _{1 \leq i < j \leq N} \, (z_{i} - z_{j})^{a_{ij}} . \]
\item Let $E (N)$ denote the set of all $N \times N$ symmetric matrices $A:=[a_{ij}]$ such that
each $a_{ij}$ is a nonnegative integer and $a_{ii} = 0$ for $1 \leq i \leq N$. For
$V := (d_{1}, \ldots , d_{N}) \in {\Bbb Z}^N$, let $E(N, V)$ be the subset of $E (N)$
consisting of all $A \in E (N)$ such that
$\rho (A)=V$. If $V = (d, d, \ldots , d)$, then $E(N, V)$ will be denoted by $E(N, d)$. Note that a member
of $E(N,d)$ may be regarded as the adjacency matrix of a $d$-regular loopless multi-graph on $N$ vertices.
\item For a positive integers $m$, $n$, define $D_{(m, n)}$ to be the $m \times n$ matrix
$[c_{ij}]$, where
\[ c_{ii} \, :=  \, \left \{ \begin{array}{ll} 0 & \mbox{if $i = j$,} \\
1 & \mbox{if $i \neq j$.} \end{array} \right . \]
By $D_{n}$, we mean the $n \times n$ matrix $D_{(n, n)}$.
\item The {\it discriminant} $\Delta  (z) \in {\Bbb Q}[z_{1}, \dots, z_{N}]$ is
defined to be $\delta (z, 2 D_{N})$, {\it i.e.},
\[ \Delta (z)\, := \, \prod_{1 \leq i < j \leq N} (z_{i} - z_{j})^2 .\]
\item Let $m$ be a positive integer and let $\sigma \in S_{m}$ denote
the $m$-cycle $(1 2 \cdots m)$. Given an ordered $m$-tuple
\[ {\frak a} \, := \, (a (1), \dots , a (m)), \] let $\mbox{cirmat} ({\frak a}) $
denote the $m \times m$ {\it circulant} matrix $[c_{ij} ]$ determined by
${\frak a}$, {\it i.e.}, for $1 \leq i, j \leq m$, let
\[ c_{ij} \, := \,a (\sigma ^{1 - i} (j)) . \]
\item Let $m$, $n$ be positive integers such that $m n = N$. Let $a$, $c$ be indeterminates.
Let ${\frak u} := (u (1), \dots , u (m))$ be defined by
\[u (i) \, := \, \left \{ \begin{array}{ll} 2 c & \mbox{if $1 \leq i \leq \frac{m-1}{2}$,} \\
0 & \mbox{otherwise.} \end{array} \right . \]
Let $M_{0} (m, n, a, c)$ be the $N \times N$ symmetric matrix defined as an
$n \times n$ block-matrix $[M_{ij}]$, where, for $1 \leq i, j\leq n$,
\[ M_{ij} \, := \, \left \{\begin{array}{ll} 2 a D_{m} & \mbox{if $i =j$,} \\
\mbox{cirmat} ({\frak u}) & \mbox{if $i < j$,} \\
\mbox{cirmat} ({\frak u})^{T} & \mbox{if $i > j$.} \end{array} \right . \]
\end{enumerate}

\noindent \underline {\bf Examples}:
\[ M_{0} (3, 2, a, c) \, = \, \mat{6}{0 & 2 a & 2 a & 2 c & 0 & 0 \\
2 a & 0 & 2 a & 0 &  2 c & 0 \\ 2 a & 2 a & 0 & 0 & 0 & 2 c \\
2 c & 0 & 0 & 0 & 2 a & 2 a \\ 0 & 2 c & 0 & 2 a & 0 & 2 a \\
0 & 0 & 2 c & 2 a & 2 a & 0}  \]
and  \[ M_{0} (5, 2, a, c) \,
= \, \mat{2}{ 2 a D_{5} & U  \\ U^{T} & 2 a D_{5}} , \]
where
\[ U  \, := \, \mat{5}{2 c & 2 c & 0 & 0 & 0 \\ 0 & 2 c & 2 c & 0 & 0 \\
0 & 0 & 2 c & 2 c & 0 \\ 0 & 0 & 0 & 2 c & 2 c  \\ 2 c & 0 & 0 & 0 & 2 c} . \] \\

\noindent \underline {\bf Theorem 2}: Let $m$, $n$, $N$ be integers such that
$2 \leq m \leq n \leq N$. Let $a$, $b$, $c$ be positive integers. Let $k$ be
a field containing ${\Bbb Q}$ and let $z_{1}, \dots, z_{N}$ be indeterminates.
As before, $z$ stands for $(z_{1}, \dots , z_{N})$.
\begin{description}
\item[(i)] Let $n$ be a positive integer and for $1 \leq i \leq n$, let
$g _{i} \in {\Bbb Q}(z_{1}, \dots , z_{N})$ be such that $g_{1} \neq 0$.
Then, $g_{1}^2 + g_{2}^2 + \cdots + g_{n}^2 \neq 0$. In particular, given a
$ 0 \neq g \in {\Bbb Q}(z_{1}, \dots , z_{N})$ and a nonempty subset
$S \subseteq S_{N}$, we have
\[ \sum _{\sigma \in S} g (z_{\sigma (1)}, \dots , z_{\sigma (N)})^{2} \, \neq  \, 0 . \]
\item[(ii)] Let $m, n, a, c$ be positive integers such that
$3 \leq m \leq n m = N$ and $m$ is odd. Then letting $M_{0} := M_{0}(m, n, a, c)$,
 we have
\[ M_{0} \in E (N, (2 a + c n - c)(m - 1)) \] and
\[ Symm_{N} ( \delta (z, M_{0}) ) \,  \neq  \, 0 . \]
\end{description}

\noindent \underline{\bf Proof}: To prove (i), let $h := g_{1}^2 + g_{2}^2 + \cdots + g_{n}^2$.
For $1 \leq  i \leq n$, let $p_{i} , q_{i} \in {\Bbb Q}[z_{1}, \dots , z_{N}]$
be polynomials such that $g_{i} q_{i} = p_{i}$ and $q_{i} \neq 0$. Note that,
$g_{1} \neq 0$ implies $p_{1} \neq 0$. Now since $f:= p_{1} q_{1} q_{2} \cdots q_{n}$ is a nonzero
polynomial, there exists $(a_{1}, \dots , a_{N}) \in {\Bbb Q}^{N}$ such that
$f(a_{1}, \dots , a_{N}) \neq 0$. Fix such $(a_{1}, \dots , a_{N})$ and let
$c_{i} := g_{i} (a_{1}, \dots , a_{N})$ for $1 \leq i \leq n$. Then each $c_{i}$ is a
rational number and $c_{1} \neq 0$. Since $c_{1}^2 > 0$ and $(c_{2}^2 + \cdots + c_{n}^2) \geq 0$,
we have $h (a_{1}, \dots , a_{N}) > 0$. In particular, $h  \neq 0$. This proves (i).

To prove (ii), let $M_{ij}$ denote the $ij$-th $m \times m$ block of $M_{0}$ (as in the definition
of $M_{0}$). If $1 \leq i < j \leq n$, then $M_{ij}$ being a circulant matrix and $m$ being odd, each
row-sum as well as each column-sum of $M_{ij}$ is exactly $ c (m - 1)$. Now it is easily verified that
$M_{0}$ is a member of $E (N, (2 a + c n - c)(m - 1))$. Since each entry of $M_{0}$ is a nonnegative
even integer, there exists a nonzero polynomial $g  \in k[z_{1}, \dots , z_{N}]$ such that
\[ Symm _{N} (\delta (z, M_{0})) \,  =  \, \sum _{\sigma  \in S_{N}} \sigma (g (z_{1}, \dots , z_{N}))^2 . \]
Therefore, (ii) follows from (i). $\Box$ \\

\noindent \underline {\bf Remarks}:
\begin{enumerate}
\item Here is an open problem related to (i) of the above theorem. Under what conditions on
$E \in E(N)$ does there exist an $a :=(a_{1}, \dots , a_{N})$ in ${\Bbb R}^{N}$ such
that $\delta (\sigma (a), E)$ is positive for all $\sigma \in S_{N}$?
\item In the direction opposite to (i), we may ask: for what $E \in E(N)$,
if any, is $Symm _{N} (\delta (z, E))$ a sum of squares of real polynomials?
For a graph-theoretic investigation of this problem, the reader is referred to~\cite{AS}. \\
\end{enumerate}

\noindent \underline {\bf Definitions}: Let $N$ be an integer such that
$2 \leq N$. As before, let $k$ be a field containing ${\Bbb Q}$, let
$z_{1}, \dots, z_{N}$ be indeterminates and let $z$ stand for
$(z_{1}, \dots , z_{N})$. In what follows, ${\Bbb N}$ denotes the set
of nonnegative integers.
\begin{enumerate}
\item Given a subset $B$ of $\{1, 2, \dots , N \}$, let
\[ \pi (B) \, := \, \{ (i, j) \in B \times B \, \mid \, i < j \}.  \]
The set $\pi (B) $ is tacitly identified with the set of all $2$-element
subsets of the set $B$, {\it i.e.},
\[ \pi (B) \, = \, \{ \{i, j \} \, \mid \, i, j \in B\;\;\mbox{and $i \neq j$} \} .\]
By $\pi [N]$, we mean the set $\pi (\{1, \dots , N \})$.
\item Given a subset $C \subseteq \pi [N]$ and a function
$\varepsilon : C \rightarrow {\Bbb N}$, the image of $(i, j) \in C$
via $\varepsilon $ is denoted by $\varepsilon (i, j)$.
A nonnegative integer $w$ is identified with the constant
function $C \rightarrow {\Bbb N}$ that maps each member of $C$ to $w$.
\item For a subset $C \subseteq \pi [N]$ and a function
$\varepsilon : C \rightarrow {\Bbb N}$, define
\[ v(z, C, \varepsilon) \, := \, \prod_{(i, j) \in C} \, (z_{i} - z_{j})^{\varepsilon (i, j)} .\]
By convention, $v (z, \emptyset, \varepsilon) = 1$.
\item Let $p$ be a positive integer and for $1 \leq r \leq p$, let
\[ S_{r} \, := \, \{ i \mid n_{r-1} + 1 \leq i \leq n_{r} \}, \]
where ${\frak n} : \, 0 = n_{0} < n_{1} < \cdots < n_{p} = N$ is a sequence of integers.
Let $a$, $c$ be indeterminates.
Let $M ({\frak n}, a, c)$ denote the $N \times N$ symmetric matrix $[u (i, j)]$
whose upper-triangular entries are defined by
\[ u (i, j) \, := \, \left \{ \begin{array}{ll} 2 a &
\mbox{if $(i, j) \in \pi (S_{r})$,} \\ 0 &
\mbox{if $(i, j) = (\epsilon + n_{r-1}, \epsilon + n_{s-1}) \in S_{r} \times S_{s}$
and $r < s$, } \\
c & \mbox{otherwise.} \end{array} \right . \]
\item Let $m$, $n$ be positive integers such that $m \geq 2$. Let $a$, $c$ be indeterminates.
Let $M (m, n, a, c)$ denote the $(m n) \times (m n)$ symmetric matrix $[a (i, j)]$ whose
entries are defined as follows: assuming $(i, j) = (l_{1} m + r_{1}, l_{2} m + r_{2})$,
where $0 \leq l_{1}, l_{2} \leq n - 1$ and $1 \leq r_{1} , r_{2} \leq m$,
\[ a (i, j) \, := \, \left \{ \begin{array}{ll} 2 a & \mbox{if $l_{1} = l_{2}$ and $r_{1} \neq r_{2}$,} \\ 0 &
\mbox{if $r_{1} = r_{2}$, } \\ c & \mbox{otherwise.} \end{array} \right . \] \\
\end{enumerate}

\noindent \underline {\bf Remarks}:
\begin{enumerate}
\item There is an obvious bijective correspondence between functions
$\varepsilon$ from  $\pi [N]$ to ${\Bbb N}$ and $N \times N$ symmetric matrices
$[a_{ij}]$ having  \[ a_{ii} \, =  \, 0 \;\;\;\; \mbox{ for $1 \leq i \leq N$, } \]
given by the prescription
\[ a_{ij} \,= \,a_{ji}\, = \,\varepsilon (i, j)\;\;\;\mbox{for $1 \leq i < j \leq N$.} \]
\item Given an integer sequence
\[ {\frak n} : \, 0 = n_{0} < n_{1} < \cdots < n_{p} = N , \]
and integers $a$, $c$ as in the above definition, $M ({\frak n}, a, c)$
is realized as a $p \times p$ block-matrix $[M_{ij}]$, where
\[ M_{rr} \, = \,  2 a D_{n_{r} - n_{r-1}} \;\;\;\;\mbox{for $1 \leq r \leq p$} \]
and if $1 \leq r \neq s \leq p$, then
\[ M_{rs} \, := \, c D_{(n_{r} - n_{r-1}, n_{s} - n_{s-1})} . \]
Likewise, $M (m, n, a, c)$ is seen to be an $n \times n$ block-matrix $[M_{ij}]$, where
\[ M_{rr} \, = \,  2 a D_{m} \;\;\;\;\mbox{for $1 \leq r \leq n$}, \]
and if $1 \leq r \neq s \leq p$, then $M_{rs} := c D_{m}$.
Some concrete examples follow. \\
\end{enumerate}

\noindent \underline {\bf Examples}:
\begin{enumerate}
\item Let $N = 4$, $p = 3$ and ${\frak n} := 0 < 1 < 3 < 4$. Then we have $M ({\frak n}, a, c)$ is
\[ M ({\frak n}, a, c) \, = \, \mat{4}{0 & 0 & c & 0 \\ 0 & 0 & 2 a & 0 \\
c & 2 a & 0 & c \\ 0 & 0 & c & 0}. \]
Note that \[ M ({\frak n}, a, c) \, = \, \mat{4}{0 & 1 &0 & 0 \\ 0 & 0 & 1 & 0 \\
1 & 0 & 0 & 0 \\ 0 & 0 & 0 & 1} \mat{4}{0 & c & 2 a & c \\ c & 0 & 0 & 0 \\
2 a & 0 & 0 & 0 \\ c & 0 & 0 & 0} \mat{4}{0 & 0 & 1 & 0 \\ 1 & 0 & 0 & 0 \\
0 & 1 & 0 & 0 \\ 0 & 0 & 0 & 1}  .\]
\item For $N = 8$ and $m = 4$, we have
\[ M (4, 2, a, c) \, := \,
\mat{8}{0 & 2 a & 2 a & 2 a & 0 & c & c & c \\ 2 a & 0 & 2 a & 2 a & c & 0 & c & c \\
2 a & 2 a & 0 & 2 a & c & c & 0 & c \\ 2 a & 2 a & 2 a & 0 & c & c & c & 0  \\
0 & c & c & c & 0 & 2 a & 2 a & 2 a \\ c & 0 & c & c & 2 a & 0 & 2 a & 2 a \\
c & c & 0 & c & 2 a & 2 a & 0 & 2 a \\ c & c & c & 0 & 2 a & 2 a & 2 a & 0}. \] \\
\end{enumerate}

\noindent \underline {\bf Theorem 3}: Let $q, m_{1}, \dots m_{q}, N$ be positive integers
such that $q \geq 2$,
\[ m_{1} \leq m_{2} \leq \cdots \leq m_{q} \;\;\;\mbox{and} \;\;\;\;
 \sum_{i=1}^{q} m_{i} \, = \, N. \]
Let $m_{0} = 0$. For $1 \leq r \leq q$, define
\[ A_{r} \, := \, \{ i + \sum _{j = 0}^{r - 1} m_{j}  \, \mid \, 1 \leq i \leq m_{r} \} . \]
Suppose $\varepsilon : \pi [N] \rightarrow {\Bbb  N}$ is a function such that
\[ (1)\;\;\;\;\;\;\; \varepsilon (i, j) \,= \, 0\;\;\;\mbox{if and only if
$(i, j) \in \pi (A_{r})$ with $1 \leq r \leq q$} \]
and $\varepsilon$ satisfies at least one of the following conditions.
\begin{description}
\item[(2)] For $1 \leq r < s \leq q$, there is a nonnegative integer $b (m_{r}, m_{s})$
(depending only on $(m_{r}, m_{s})$) such that $b (m_{r}, m_{s})$ is even if $m_{r} = m_{s}$ and
\[ \sum _{(i, j) \in A_{r} \times A_{s}}  \varepsilon (i, j) \, = \, b (m_{r}, m_{s}). \]
\item[(3)] For $1 \leq r < s \leq q$,
\[ \sum _{(i, j) \in A_{r} \times A_{s}}  \varepsilon (i, j) \;\;\;\;\mbox{is an even integer.} \]
\end{description}
Then the following holds.
\begin{description}
\item[(i)] Letting $\mu (z, \varepsilon)\, := \, v (z, \pi[N], \varepsilon)$, we have
\[ Symm_{N} ( \mu (z, \varepsilon) ) \, \neq \, 0. \]
\item[(ii)] Suppose $M := M({\frak n}, a, c)$, where $a, c, p$ are positive integers,
and \[ {\frak n} : \, 0 = n_{0} < n_{1} < \cdots < n_{p} = N \] is a sequence
of integers. For $1 \leq r \leq p$ and $n_{r-1} + 1 \leq i \leq n_{r}$, let $\chi (i)$
denote the number of $1 \leq s \leq p$ with $(n_{s} - n_{s -1}) < (i - n_{r - 1})$. Let
$W := (w_{1}, \dots , w_{N})$ be such that if $n_{r-1} + 1 \leq i \leq n_{r}$, then
\[ w_{i} \, :=  \, (2 a - c) (n_{r} - n_{r-1} - 1) + c (N + \chi (i) - p). \]
Then we have $M \in E (N, W)$ and
\[ Symm_{N} ( \delta (z, M) )\,  \neq  \, 0 . \]
\end{description}

\noindent \underline{\bf Proof}: At the outset, we observe that each
$\sigma \in S_{N}$ can be naturally viewed as a permutation of $\pi [N]$
by letting $\sigma (i, j) \, := \, \{ \sigma (i), \sigma (j) \}$, {\it i.e.},
for $(i, j) \in \pi [N]$,
\[\sigma (i, j) \, := \,
\left \{\begin{array}{ll} (\sigma (i), \sigma (j)) &
\mbox{if $\sigma (i) < \sigma (j)$,} \\ (\sigma (j), \sigma (i)) &
\mbox{if $\sigma (j) < \sigma (i)$.} \end{array} \right . \]
In this manner $S_{N}$ is regarded as a subgroup of the group of permutations
of $\pi [N]$. For $\sigma \in S_{N}$ and $1 \leq r \leq q$, define
\[ B_{r} (\sigma)\, := \, \sigma ^{-1} (A_{r}) \, = \,
\{ i \, \mid \, 1 \leq i \leq N \;\;\mbox{and}\;\;\sigma (i) \in A_{r} \}. \]
Clearly, sets $B_{1} (\sigma), \dots , B_{q} (\sigma)$ partition $\{1, \dots , N \}$ and $B_{i}$
has cardinality $m_{i}$ for all $1 \leq i \leq q$.

Define
\[ \pi \, := \, \pi [N] \setminus \bigcup _{i = 1}^{q} \pi (A_{i}) \, = \,
\bigcup _{1 \leq r < s \leq q} A_{r} \times A_{s}  \]
and let
\[ G \, := \, \{ \sigma \in S_{N} \, \mid \, \sigma (i, j) \in \pi
\;\;\mbox{for all $(i, j) \in \pi$} \} . \]
In view of hypothesis (1), we have $\mu(z, \varepsilon) = v(z, \pi, \varepsilon)$.
Given $\sigma \in G$ and $(i, j) \in \pi (A_{r})$ with $1 \leq r \leq q$,
clearly there is a unique $s$ with $1 \leq s \leq q$ such that $\sigma (i , j) \in \pi (A_{s})$.
Fix a $\sigma \in G$. Consider $i \in B_{r} (\sigma) \cap A_{s}$ with $1 \leq s \leq q$.
Then, for $i \neq j \in A_{s}$, we must have $\{\sigma (i), \sigma (j) \}$ in $\pi (A_{r})$
and hence $j \in B_{r} (\sigma)$. It follows that $A_{s} \subseteq B_{r} (\sigma)$. If
$1 \leq s < p \leq q$ are such that $A_{s} \cup A_{p} \subseteq B_{r} (\sigma)$, then
an $(i, j) \in A_{s} \times A_{p}$ is in $\pi$ whereas $\sigma (i, j)$ is
in $\pi (A_{r})$. This is impossible since $\sigma \in G$. Thus we have established
the following: given $r$ with $1 \leq r \leq q$ and $\sigma \in G$, there is a unique
integer $r (\sigma)$ such that $1 \leq r(\sigma) \leq q$ and $B_{r} (\sigma) = A_{r (\sigma)}$.
In other words, the image sets $\sigma (A_{1}), \dots , \sigma (A_{q})$ form a
permutation of the sets $A_{1}, \dots , A_{q}$. For $1 \leq r < s \leq q$, since
$r (\sigma) \neq s (\sigma)$, we have
\[ \pi \cap \left ( A_{r (\sigma)} \times A_{s (\sigma)} \right ) \, \neq \, \emptyset
\;\;\;\mbox{if and only if $r (\sigma) < s (\sigma)$.} \]
Moreover,
\[ m_{r (\sigma)} \, = \, m_{r} \;\;\; \mbox{for all $1 \leq r \leq q$ and $\sigma \in G$.} \]

Now let $t, t_{1}, \ldots , t_{q},  x_{1}, \dots , x_{N}$ be indeterminates and let
\[ \alpha : k [z_{1}, \ldots , z_{N}] \rightarrow
k [t, t_{1}, \ldots , t_{q}, x_{1}, \dots , x_{N}] \]
be the injective $k$-homomorphism of rings defined by
\[ \alpha (z_{i}) \, := \, t x_{i} + t_{r} \;\;\;
\mbox{if $i \in A_{r}$ with $1 \leq r \leq q$.}  \]
Then, given $\sigma \in S_{N}$, $(i, j) \in \pi [N]$ and $1 \leq r, s \leq q$, we have
\[ \alpha (z_{\sigma (i)} - z_{\sigma (j)}) \, = \,
t( x_{\sigma (i)} - x_{\sigma(j)} ) + (t_{r} - t_{s}) \]
if and only if $(\sigma (i), \sigma (j)) \in A_{r} \times A_{s}$.

Let $x$ stand for $(x_{1}, \dots , x_{N})$ and $T$ stand for $(t_{1}, \dots , t_{q})$.
Given $f \in k [t, T, X]$, by the {\it $x$-degree} (resp. {\it $T$-degree}) of $f$,
we mean the total degree of $f$ in the indeterminates $x_{1}, \dots , x_{N}$
(resp. $t_{1}, \dots , t_{q}$). Now fix a $\sigma \in G$ and consider
\[ V_{\sigma} (x, t, T)\, := \, \alpha ( \sigma (v (z, \pi, \varepsilon ))). \]
For an ordered pair $(i, j)$ with $1 \leq i, j \leq q$, set
\[ A (\sigma, i, j) \, := \, \pi \cap ( A_{i (\sigma)} \times A_{j (\sigma)}) .\]
It is straightforward to verify that $V_{\sigma} (x, 0, T)$ is
\[ \prod _{1 \leq r < s \leq q} \left ( \prod _{(i, j) \in A (\sigma, r, s)} (t_{r} - t_{s})^{\varepsilon (i, j)}
  \cdot \prod _{(i, j) \in A (\sigma, s, r)} (t_{s} - t_{r})^{\varepsilon (i, j)} \right ).   \]
Consider an ordered pair $(r, s)$ with $1 \leq r < s \leq q$. If $s (\sigma) < r (\sigma)$, then
\[ m_{s} = m_{s (\sigma)} \leq m_{r(\sigma)} = m_{r} \;\;\;\mbox{and hence
$m_{s} = m_{s (\sigma)}  = m_{r(\sigma)} = m_{r}$.} \]
Combining this observation with condition (2), we deduce that
\[\sum _{(i, j) \in A_{s (\sigma)} \times A_{r (\sigma)}} \varepsilon (i, j) \, = \,
\left \{\begin{array}{ll} 0 & \mbox{if $r (\sigma) < s (\sigma)$,} \\
b (m_{r}, m_{s}) & \mbox{if $s (\sigma) < r (\sigma)$,} \end{array} \right . \]
is an even integer and hence
\[ V_{\sigma} (x, 0, T)\, := \, \prod _{1 \leq r < s \leq q} (t_{r} - t_{s})^{ b(m_{r}, m_{s})} .\]
On the other hand, if condition (3) holds, then we simply note that there is a nonzero homogeneous polynomial
$g_{\sigma} \in {\Bbb Q}[t_{1}, \dots , t_{q}]$ such that $V_{\sigma} (x, 0, T) = g_{\sigma}^{2}$.
In any case, the $t$-order of $V_{\sigma} (x, 0, T)$ is $0$, {\it i.e.}, $V_{\sigma} (x, t, T)$ is not
a multiple of $t$, and the $T$-degree of $V_{\sigma} (x, 0, T)$ is
\[ d \, := \, \sum _{(i, j) \in \pi} \varepsilon (i, j). \]
Define
\[ \gamma\, := \, \sum _{\sigma \in G} \, \sigma ( \mu (z, \varepsilon) ) \;\;\;\mbox{and}
\;\;\; V (x, t, T) \, := \, \sum _{\sigma \in G}\, V_{\sigma} (x, t, T) . \]
Then $\alpha (\gamma) =  V (x, t, T)$. If (2) holds, then letting $|G|$ denote the cardinality of $G$,
we have
\[ V (x, 0, T) \, = \, |G|\,\prod _{1 \leq r < s \leq q} (t_{r} - t_{s})^{b (m_{r}, m_{s})} \]
and hence $V (x, 0, T) \neq 0$. On the other hand, if (3) holds, then we have
\[ V(x, 0, T) \, = \, \sum _{\sigma \in G}\, g_{\sigma}^{2}, \]
which is necessarily nonzero in view of (i) of Theorem 2. Now it is clear that
$\alpha (\gamma) \neq 0$, the $t$-order of $\alpha (\gamma)$ is $0$ and the $T$-degree of
$\alpha (\gamma)$ is $d$.

Next, for $\sigma \in S_{N}$, let
\[ R (\sigma) \, := \, \bigcup _{1 \leq r \leq q} \pi (B_{r} (\sigma)) . \]
Observe that $ \pi \cap R (\sigma) =  \emptyset $ if and only if $\sigma \in G$ and
\[ \alpha (z_{\sigma (i)} - z_{\sigma (j)}) \, = \, t ( x_{\sigma (i)} - x_{\sigma (j)} ) \]
if and only if $(i, j) \in R (\sigma)$; otherwise, we have
\[ \alpha (z_{\sigma (i)} - z_{\sigma (j)}) \, = \,t( x_{\sigma(i)} - x_{\sigma (j)} ) + (t_{r} - t_{s}) \]
for some $1 \leq r, s \leq q$ with $r \neq s$.

Now fix a $\sigma \in S_{N} \setminus G$. Then $\mu (z, \varepsilon)$ is readily seen to be
the product of $v(z, R (\sigma), \varepsilon)$ and $v (z, \pi [N] \setminus R (\sigma), \varepsilon)$.
Moreover,
\[v(z, R (\sigma), \varepsilon) \, = \, v (z, \pi \cap R (\sigma), \varepsilon), \]
and similarly
\[v(z, \pi [N] \setminus R (\sigma), \varepsilon) \, = \, v (z, \pi \setminus R (\sigma), \varepsilon). \]
Define
\[ \begin{array}{l} \lambda (\sigma) \, := \, \sum _{(i, j) \in \pi \cap R (\sigma)} \varepsilon (i, j)
\;\;\;\mbox{and} \\
d (\sigma) \, := \, \sum _{(i, j) \in \pi \setminus R (\sigma)} \varepsilon (i, j) . \end{array} \]
Then $d (\sigma) \, = \, d - \lambda (\sigma)$. From our choice of $\sigma$ and hypothesis (1), it follows
that $\lambda (\sigma) \geq 1$ and hence $d (\sigma) < d $. Let
\[\begin{array}{l} P_{\sigma} (x, t, T) \, := \, \alpha (\sigma (v (z, \pi \cap R (\sigma), \varepsilon))), \\
Q_{\sigma} (x, t, T) \,:= \,\alpha (\sigma (v (z, \pi \setminus R (\sigma), \varepsilon))). \end{array}  \]
Observe that
\[ P_{\sigma} (x, t, T) \, = \, t^{\lambda (\sigma)} \cdot
\prod _{(i, j) \in \pi \cap R (\sigma)} (x_{\sigma (i)} - x_{\sigma (j)})^{\varepsilon (i, j)} \]
and $Q_{\sigma}(x, 0, T)$ is a nonzero $T$-homogeneous polynomial of $T$-degree $d (\sigma)$. In particular,
the $t$-order of the product $P_{\sigma} (x, t, T) \cdot Q_{\sigma} (x, t, T)$ is exactly
$\lambda (\sigma)$. Consequently, for each $\sigma \in S_{N} \setminus G$, the $t$-order of
$\alpha (\sigma (\mu (z, \varepsilon)))$ is positive. Define
\[ \psi \, := \, \sum _{\sigma \in S_{N} \setminus G} \, \sigma ( \mu (z, \varepsilon) ). \]
Then the $t$-order of $\alpha (\psi)$ is strictly positive and hence the $t$-order of
$\alpha (\gamma) + \alpha (\psi)$ is zero; in particular $\alpha (\gamma) + \alpha (\psi) \neq 0$.
Since $Symm_{N} (\mu (z, \varepsilon)) = \gamma + \psi$ and
\[ \alpha ( Symm_{N} (\mu (z, \varepsilon))) \, = \, \alpha (\gamma) + \alpha (\psi) \, \neq \, 0,\]
our assertion (i) holds.

Consider $M$ as in (ii). The block-format description of $M$ presented in the remarks following
the definitions above allows a straightforward verification that the matrix $M$ belongs to $E (N, W)$.
To prove the rest of (ii), let
\[ q \,:= \, \mbox{max}\,\{n_{i} - n_{i-1} \mid 1 \leq i \leq p \}\;\;\; \mbox{and} \;\;\;
\Gamma (r) \, := \, \{1 \leq i \leq p \mid (n_{i} - n_{i-1}) \geq r \} \]
for $1 \leq r \leq q$. Let $\Omega (r)$ denote the cardinality of $\Gamma (r)$ and define
\[ K_{r} \, := \, \{ r + n_{i - 1} \, \mid \, i \in \Gamma (r)\}. \]
Then $K_{r}$ has cardinality $\Omega (r) \geq 1$ for $1 \leq r \leq q$, and the sets
$K_{1}, \dots , K_{q}$ partition $\{1, \dots , N \}$.
Note that letting $M = [u (i, j)]$, we have $u (i, j) = 0$ if and only if $(i, j) \in K_{r} \times K_{r}$
for some $r$ with $1 \leq r \leq q$. Likewise, $u (i, j) =  2 a$ if and only if $(i, j) \in K_{r} \times K_{s}$,
$i - r = j - s$ and $1 \leq r \neq s \leq q$. Furthermore, it is easy to verify that
\[ \sum _{(i, j) \in K_{r} \times K_{s}} u (i, j) \, = \,
c \Omega (r) \Omega (s) + (2 a - c) \cdot \mbox{min}\,\{\Omega (r), \Omega (s) \}. \]
Let $\tau$ be a permutation of $\{ 1, \dots , q \}$ such that $\Omega (\tau (i)) \leq \Omega (\tau (j))$
for $1 \leq i < j \leq q$. Let $m_{0} := 0$ and $m_{i} := \Omega (\tau (i))$ for $1 \leq i \leq q$. Then
we have $1 \leq m_{1} \leq m_{2} \leq \cdots \leq m_{q}$ and $m_{1} + m_{2} + \cdots + m_{q} = N$. Let
$\theta \in S_{N}$ be a permutation such that
\[ A_{r} \, := \, \{ \theta (i) \mid i \in K_{\tau (r)} \} \, = \,
 \{ i + \sum _{j = 0}^{r - 1} m_{j}  \, \mid \, 1 \leq i \leq m_{r} \} . \]

Existence of $\theta$ is assured by the obvious fact that $K_{\tau (1)},  K_{\tau (2)}, \dots , K_{\tau (q)}$
partition $\{1, 2, \dots , N \}$. Observe that $q = 1$ if and only if $N = p$ if and only if $M = 0$.
Since $\delta (z, 0) = 1$ leads to $Symm _{N} (\delta (z, 0)) = N! \neq 0$, we henceforth assume that $q \geq 2$.
Now, let $\varepsilon : \pi [N] \rightarrow {\Bbb N}$ be defined by
\[ \varepsilon (i, j) \, := \, u (\theta ^{-1} (i), \theta ^{-1} (j)) \;\;\;
\mbox{ for $(i, j) \in \pi [N]$.} \]
Then $\varepsilon (i, j) = 0$ if and only if $(i, j) \in \pi (A_{r})$ for $1 \leq r \leq q$.
Furthermore, for $1 \leq r < s \leq q$, since
\[ \sum _{(i, j) \in A_{r} \times A_{s}} \, \varepsilon (i, j) \, = \,
\sum _{(i, j) \in K_{\tau (r)} \times K_{\tau (s)}} \, u (i, j) , \]
we get
\[ \sum _{(i, j) \in A_{r} \times A_{s}} \, \varepsilon (i, j) \, = \,
c m_{r} m_{s} + (2 a - c) m_{r} .\]
So, for $1 \leq r < s \leq q$, by letting $b (m_{r}, m_{s})$ to be the number on the right of
the above equality, $\varepsilon$ is seen to satisfy the condition (2). Finally, note that
\[ \theta (\delta (z, M)) \, = \,
\prod _{(i, j) \in \pi [N]} (z_{\theta (i)} - z_{\theta (j)})^{u (i, j)} \, = \,
\pm \prod _{(i, j) \in \pi [N]} (z_{i} - z_{j})^{\varepsilon (i, j)}  \]
and the right-most product in the above equation is $\pm \mu (z, \varepsilon)$. Since
\[Symm _{N} (\delta (z, M)) \, = \, Symm _{N} (\theta (\delta (z, M))
\, = \pm \, Symm _{N} (\mu (z, \varepsilon))  \]
and $Symm_{N} (\mu (z, \varepsilon)) \neq 0$ by (i), assertion (ii) stands verified. $\Box$ \\

\noindent \underline {\bf Corollary}: Let $m$, $n$ be positive integers such that
$m \geq 2$ and $N = m n$. For $1 \leq r < s \leq m$, define
\[T (r, s) \, := \, \{ \{l_{1} m + r,\, l_{2} + s \} \, \mid \, 0 \leq l_{1}, l_{2} \leq n - 1 \} . \]
As declared before, we identify $T(r, s)$ as a subset of $\pi [N]$.
\begin{description}
\item[(i)] Let $a : \pi[N] \rightarrow {\Bbb N}$ be a function such that
\[a (i, j) \, =  \, 0 \;\;\; \mbox{if and only if $i \equiv j \,\mbox{mod}\, m$}, \]
and for $1 \leq r < s \leq m$,
\[ \sum _{(i, j) \in T (r, s)} a (i, j) \;\;\;\;\mbox{is an even integer.} \]
Then, we have
\[ Symm _{N} ( v(z, \pi [N], a) ) \, \neq \, 0 . \]
\item[(ii)] Let $a$, $c$ be positive integers. Then, letting $M := M(m, n, a, c)$, we have
\[ M \in E (N,  (2 a + c n - c)(m - 1)) \] and
\[ Symm_{N} ( \delta (z, M) ) \,  \neq  \, 0 . \] \\
\end{description}

\noindent \underline{\bf Proof}: Let $\theta \in S_{N}$ be the permutation defined by
\[ \theta (i) \, := \, l + 1 + (r - 1) n  \;\;\; \mbox{if $i =  l m + r$ with
$0 \leq l \leq n - 1$ and $1 \leq r \leq m$.} \]
Now specializing $q$ to $m$ and $m_{i}$ to $n$ for $1 \leq i \leq q$, we have
\[ A_{r} \, = \, \{(r -1) n + l + 1 \, \mid \, 0 \leq l \leq n - 1 \} \, = \,
\{ \theta ( l m + r ) \, \mid \, 0 \leq l \leq n - 1 \} \]
for $1 \leq r \leq m$. Next, for our function $a$ as in (i), let $\varepsilon$ be the
function defined as follows:
\[ \varepsilon (i, j) \, := \, a ( \{\theta ^{-1} (i), \theta ^{-1} (j) \} ) \;\;\;
\mbox{ for all $(i, j) \in \pi [N]$.} \]
Then $\varepsilon (i, j)$ is a nonnegative integer for all
$(i, j) \in \pi [N]$ and $\varepsilon (i, j) = 0$ if and only if $(i, j) \in \pi (A_{r})$
with $1 \leq r \leq m$. Furthermore, given $1 \leq r < s \leq m$ and
$(i, j) \in A_{r} \times A_{s}$, letting
\[i = (r - 1) n + l_{1} + 1 \;\;\;\; j = (s - 1) n + l_{2} + 1 \]
with $0 \leq l_{1}, l_{2} \leq n - 1$, we have
\[ a ( \{ \theta ^{-1} (i), \theta ^{-1} (j) \}) \, = \, a ( \{ l_{1} m + r, l_{2} m + s \} ) \]
and hence
\[ \sum _{(i, j) \in A_{r} \times A_{s}} \, \varepsilon (i, j) \, = \, \sum _{(i, j) \in T (r, s)} a (i, j).  \]
Since the right side of the above equation is an even integer by our hypothesis, $\varepsilon$ clearly satisfies
conditions (1) and  (3) of the above theorem. Thus assertion (i) follows from (i) of the above theorem.

Let $M$ be as in (ii). Using the block-format description presented in the remarks following the above
definitions,  it is easily seen that $M$ is in $E (N,  (2 a + c n - c)(m - 1))$. Consider the integer-sequence
\[ {\frak n} \, := \, \{i m \mid 0 \leq i \leq n \} \, = \, 0 < m < 2 m < \cdots < i m < \cdots < n m = N .\]
Then it is easy to see that $M = M({\frak n}, a, c)$ and hence our assertion
follows from (ii) of the above theorem. Of course, (ii) can also be derived from (i); the details
of this derivation are left to the reader. $\Box$ \\

\noindent \underline {\bf Remarks}:
\begin{enumerate}
\item The function $\varepsilon$ of the above theorem satisfies (2) provided
the sum
\[ \sum _{(i, j) \in A_{r} \times A_{s}}  \varepsilon (i, j) \]
is an even integer whenever $m_{r} = m_{s}$ with $1 \leq r < s \leq q$ and
\[ \sum _{(i, j) \in A_{r_{1}} \times A_{s_{1}}}  \varepsilon (i, j) \, = \,
\sum _{(i, j) \in A_{r_{2}} \times A_{s_{2}}}  \varepsilon (i, j) \]
whenever $1 \leq r_{1} < s_{1} \leq q$, $1 \leq r_{2} < s_{2} \leq q$ are
such that \[ (m_{r_{1}}, m_{s_{1}}) = (m_{r_{2}}, m_{s_{2}}). \] In particular,
observe that if $m_{1} < m_{2} < \cdots < m_{q}$, then the condition (2) is
vacuously satisfied.
\item Suppose $m_{1} \leq \cdots \leq m_{q}$ and  $\varepsilon$
are as in the above theorem. Also assume that $\varepsilon$ satisfies condition (1). Then,
it is of independent interest to find a condition on $\varepsilon$, that is a simultaneous
generalization of (2), (3) and ensures nonzero symmetrization of $\mu (z, \varepsilon)$.
It is most likely that such a condition will also ensure nonzero
symmetrization of the reciprocal of $\mu (z, \varepsilon)$.  \\
\end{enumerate}

\noindent \underline {\bf Examples}: Consider the $6 \times 6$ block-matrices
\[E_{1} \, := \, \mat{2}{0 & C_{1} \\ C_{1}^{T} & 0} \;\;\;\mbox{and} \;\;\;
 E_{2} \, := \, \mat{2}{0 & C_{2} \\ C_{2}^{T} & 0}, \]
where
\[C_{1} \, := \, \mat{3}{3 & 3 & 3 \\ 3 & 4 & 3 \\ 3 & 3 & 4}\;\;\; \mbox{and} \;\;\;
C_{2} \, := \, \mat{3}{3 & 3 & 3 \\ 3 & 3 & 4 \\ 3 & 3 & 4} . \]
Then, $Symm_{6} ( \delta (z, E_{1})) = 0$ and $Symm_{6} ( \delta (z, E_{2})) \neq 0$.
This hints at the subtle nature of the difficulties involved in generalizing Theorem 3. \\

\section{Applications to the IQL state with $\nu  < 1 / 2$}

In this section, we apply the theorems of the previous in constructing correlation functions
$G(z_{1}, \dots , z_{N})$ for a system of $N$ interacting Fermions corresponding to the given
value of the filling factor $\nu$. We also record here the statistics for correlation frequency within the various pairs of Fermions in the associated correlation diagram, which might prove useful for certain energy computations (see, e.g., \cite{Quinn09}).  Recall that the trial wave function for such a system is
the product $F(z_{1}, \dots , z_{N}) G(z_{1}, \dots , z_{N})$, where
\[ F(z_{1}, \dots , z_{N}) \, := \, \prod_{1 \leq i < j \leq N} (z_{i} - z_{j})  \]
is alternating and $G(z_{1}, \dots , z_{N})$ is symmetric in $z_{1}, \dots , z_{N}$. \\

Let $N$, $n$ be positive integers such that $N \geq 3$ and $N =  m n$ for an integer $m \geq 2$.
Let $\nu$ be a rational number of the form $n / (2 p n \pm 1)$, where $p$ is a positive integer
such that $\nu < {^1/_2}$, {\it i.e.}, either $p \geq 2$ or $\nu \neq n /(2 n - 1)$.
We construct correlation functions for configurations of $N$ Fermions in the IQL state with filling
factor $\nu$. Note that the  $G(z_{1}, \dots , z_{N})$ we construct is a nonzero homogeneous polynomial having
(mandated) total degree
\[\kappa _{G} \, := \, N l - \frac{N (N - 1)}{2} \, = \, \frac{1}{2} N [ (2 p - 1) n \pm 1 ] ( m - 1).\]
Furthermore, $G(z_{1}, \dots , z_{N})$ is obtained by symmetrizing a suitable $\delta (z, E)$, where the matrix $E$ is in
$E(N, 2 l - N + 1)$. Thus, our $G(z_{1}, \dots , z_{N})$ is a binary invariant of
type $(N, 2 l - N + 1)$ (see, e.g., \cite{KR} for definition), and in particular, its $z_{i}$-degree equals $2 l - N + 1$ for $1 \leq i \leq N$.
It is interesting to observe that for all $\nu < {^1/_2}$, the corresponding number $\kappa _{G}$ is an even
integer. Consequently, as a bonus, $G(z_{1}, \dots , z_{N})$ satisfies the additional property:
\[ G(-z_{1}, \dots , -z_{N}) \, = \, G(z_{1}, \dots , z_{N}). \]
For an $E := [ \varepsilon (i, j) ] \in E(N)$ and an integer $b$, define $frq (b, E)$
({\it frequency of $b$ in $E$}) to be the cardinality of the set
\[ \{(i, j) \, \mid \, 1 \leq i < j \leq N \;\; \mbox{with} \;\; \varepsilon (i, j) = b \}. \]

\begin{enumerate}
\item Assume $\nu = n / (2 p n + 1)$. Then we have
\[ 2 l \, =  \, \nu^{-1} N - (2 p n + 1) - 1 + n. \]
Consequently,
\[2 l - N + 1 \, = \, [ (2 p - 1) n + 1 ] ( m - 1). \]
Let $G := Symm _{N} (\delta (z, E))$, where $E := M (m, n, p, 2 p - 1)$. Assertion (ii) of the
Corollary of Theorem 3 above ensures that $G(z_{1}, \dots , z_{N})$ is a nonzero polynomial.
$G(z_{1}, \dots , z_{N})$ is homogeneous of total degree
\[\kappa _{G} \, := \, N l - \frac{N (N -1)}{2} \, = \, \frac{1}{2} N [ (2 p - 1) n + 1 ] ( m - 1),\]
and its $z_{i}$-degree is $2 l - N + 1$ for $1 \leq i \leq N$. From the definition of $M (m, n, p, 2 p - 1)$,
it is clear that $suppt (E) \, =  \, \{0, 2 p - 1, 2 p \}$,
\[frq (b, E) \, = \, \left \{\begin{array}{ll} \frac{N (n -1)}{2} & \mbox{if $b = 0$,} \\ \\
 \frac{N (N - n) (n -1)}{ 2 n} & \mbox{if $b = 2 p - 1$,} \\  \\
\frac{N (N - n)}{2 n} & \mbox{if $b = 2 p$.} \end{array} \right . \]

Now, consider the special case where $m$ is an odd integer. Then $m \geq 3$ and hence we may let
$G := Symm _{N} (\delta (z, E_{0}))$, where \[ E_{0} := M_{0}(m, n, p, 2 p - 1). \] Assertion (ii) of
Theorem 2 above ensures that $G(z_{1}, \dots , z_{N})$ is a nonzero polynomial. $G(z_{1}, \dots , z_{N})$
is homogeneous of total degree
\[\kappa _{G} \, := \, N l - \frac{N (N -1)}{2} \, = \, \frac{1}{2} N [ (2 p - 1) n + 1 ] ( m - 1), \]
and its $z_{i}$-degree is $2 l - N + 1$ for $1 \leq i \leq N$. From the definition of
$M_0(m, n, p, 2 p - 1)$, it follows that $suppt (E_{0}) \, =  \, \{0, 4 p - 2, 2 p \}$,
\[frq (b, E_{0}) \, = \, \left \{\begin{array}{ll} \frac{N (N + n) (n - 1)}{4 n} & \mbox{if $b = 0$,}\\ \\
\frac{N (N - n) (n - 1)}{ 4 n} & \mbox{if $b = 4 p - 2$,} \\ \\
\frac{N (N - n)}{2 n} & \mbox{if $b = 2 p$.} \end{array} \right . \]
Observe that if $p = 1$, then $suppt (E_{0}) =\{0, 2 \}$ and
\[ frq (2, E_{0}) \, = \, \frac{N (N - n) (n + 1)}{4 n}, \;\;\; frq (0, E_{0}) \, = \, \frac{N (N + n) (n - 1)}{4 n}. \]

Lastly, consider the case where $n\geq3$ is odd and $2 p$ is an integer multiple of $n - 1$ ({\it e.g.},
$n = 3$); say $2 p = s (n - 1)$. Let  $G := Symm _{N} (\delta (z, E_{0}))$, where
$E_{0} := M_{0} (m, n, p (m - 1), s - 1)$. Again, assertion (ii) of Theorem 2 ensures that $G(z_{1}, \dots , z_{N})$
is a nonzero polynomial. $G(z_{1}, \dots , z_{N})$ is homogeneous of total degree
\[\kappa _{G} \, := \, N l - \frac{N (N -1)}{2} \, = \, \frac{1}{2} N [ (2 p - 1) n + 1 ] ( m - 1), \]
and its $z_{i}$-degree is $2 l - N + 1$ for $1 \leq i \leq N$. We have
\[ suppt (E_{0}) \, =  \, \{0, 2 p (m - 1), 2 (s - 1) \}, \]
\[ frq (b, E_{0}) \, = \, \left \{\begin{array}{ll} \frac{N (N - n) (n + 1)}{4 n} & \mbox{if $b = 0$,}\\ \\
\frac{N (n - 1)}{ 2} & \mbox{if $b = 2 p (m - 1)$,} \\ \\
\frac{N (N - n) (n - 1)}{4 n} & \mbox{if $b = 2 (s - 1)$.} \end{array} \right . \]
Observe that when $s = 1$, we have  \[ suppt (E_{0}) \, = \, \{0, 2 p (m - 1) \} \] and
\[ frq (2 p (m - 1), E_{0}) \, = \, \frac{N (n - 1)}{2}, \;\;\; frq (0, E_{0}) \, = \, \frac{N (N - n)}{2}. \]

\item Assume $\nu = n / (2 p n - 1)$. Now $l$ is the half-integer given by
\[ 2 l \, =  \, \nu^{-1} N - (2 p n - 1) - 1 + n.  \]
Also, we clearly have
\[2 l - N + 1 \, = \, [ (2 p - 1) n - 1 ] ( m - 1). \]
Let $G := Symm _{N} (\delta (z, E))$, where $E := M (m, n, p - 1, 2 p - 1)$. Assertion (ii) of the
Corollary of Theorem 3 ensures that $G(z_{1}, \dots , z_{N})$ is a nonzero polynomial.
$G(z_{1}, \dots , z_{N})$ is homogeneous of total degree
\[\kappa _{G} \, := \, N l - \frac{N (N - 1)}{2} \, = \, \frac{1}{2} N [ (2 p - 1) n + 1 ] ( m - 1),  \]
and its $z_{i}$-degree is $2 l - N + 1$ for $1 \leq i \leq N$.
From the definition of $M (m, n, p - 1, 2 p - 1)$, it is clear that $suppt (E) \, =  \, \{0, 2 p - 2, 2 p - 1 \}$,
\[ frq (b, E) \, = \, \left \{\begin{array}{ll} \frac{N (n - 1)}{2} & \mbox{if $b = 0$,}\\ \\
\frac{N (N - n)}{2 n} & \mbox{if $b = 2 p - 2$,} \\ \\
\frac{N (N - n) (n - 1)}{2 n} & \mbox{if $b = 2 p - 1$.} \end{array} \right . \]

Consider the special case where $m$ is an odd integer. Of course, we must have $m \geq 3$ and so we
let $G := Symm _{N} (\delta (z, E_{0}))$, where
\[ E_{0} := M_{0}(m, n, p - 1, 2 p - 1). \] Assertion (ii) of
Theorem 2 ensures that $G(z_{1}, \dots , z_{N})$ is a nonzero polynomial. $G(z_{1}, \dots , z_{N})$ is
homogeneous of total degree
\[\kappa _{G} \, := \, N l - \frac{N (N -1)}{2} \, = \, \frac{1}{2} N [ (2 p - 1) n + 1 ] ( m - 1),  \]
and its $z_{i}$-degree is $2 l - N + 1$ for $1 \leq i \leq N$. We have
\[ suppt (E_{0}) \, =  \, \{0, 2 p - 2, 4 p - 2 \}, \]
\[ frq (b, E_{0}) \, = \, \left \{\begin{array}{ll} \frac{N (N + n) (n - 1)}{4 n} & \mbox{if $b = 0$,}\\ \\
\frac{N (N - n)}{2 n} & \mbox{if $b = 2 p - 2$,} \\ \\
\frac{N (N - n) (n - 1)}{4 n} & \mbox{if $b = 4 p - 2$.} \end{array} \right . \]
Also, when $p = 1$, we have $suppt (E_{0}) =\{0, 2 \}$ and
\[ frq (0, E_{0}) \, = \, \frac{N (N + n) (n - 1)}{4 n}, \;\; frq (2, E_{0})\, =\, \frac{N (N - n) (n + 1)}{4 n}. \]

Lastly, consider the case where $n\geq3$ is odd and $2 (p - 1)$ is an integer multiple of $n - 1$, where $p\geq2$;
say $2 (p - 1) = s (n - 1)$ ({\it e.g.}, $\nu = 3 /(6 p - 1)$ with $p \geq 2$). Let $G := Symm _{N} (\delta (z, E_{0}))$,
where $E_{0} := M_{0} (m, n, p (m - 1), s - 1)$. Then, assertion (ii) of Theorem 2 ensures that $G(z_{1}, \dots , z_{N})$
is a nonzero polynomial. $G(z_{1}, \dots , z_{N})$ is homogeneous of total degree
\[\kappa _{G} \, := \, N l - \frac{N (N -1)}{2} \, = \, \frac{1}{2} N [ (2 p - 1) n + 1 ] ( m - 1), \]
and its $z_{i}$-degree is $2 l - N + 1$ for $1 \leq i \leq N$. We have
\[ suppt (E_{0}) \, =  \, \{0, 2 p (m - 1), 2 (s - 1) \}, \]
\[ frq (b, E_{0}) \, = \, \left \{\begin{array}{ll} \frac{N (N - n) (n + 1)}{4 n} & \mbox{if $b = 0$,}\\ \\
\frac{N (n - 1)}{2} & \mbox{if $b = 2 p (m - 1)$,} \\ \\
\frac{N (N - n) (n - 1)}{4 n} & \mbox{if $b = 2 (s - 1)$.} \end{array} \right . \]
Observe that if $s = 1$, then \[ suppt (E_{0}) =\{0, 2 p (m - 1) \} \] and
\[ frq (2 p (m - 1), E_{0}) \, = \, \frac{N (n - 1)}{2}, \;\;\;frq (0, E_{0}) \, = \, \frac{N (N - n)}{2}. \] \\
\end{enumerate}

As already observed, the Laughlin configuration, where there are exactly $2 p$ edges between each pair of vertices,
is singularly distinguished among the multitude of existent configurations for the filling factor $\nu = 1 / (2 p + 1)$
due mainly to the facts: (1) the graph-monomial of its unique correlation diagram is a symmetric polynomial, and (2) the
pair correlations are as minimal as possible. We proceed to take a broader view of this distinction. Here we propose, what we
consider to be, a fitting generalization of the Laughlin configuration to the case of $N$ Fermions in an IQL state with
filling factor $\nu = n / (2 p n \pm 1) \; < 1/2$, where $N$ is a multiple of $n$. We select a relatively small pool of
special configurations which we feel should be the prominent contributors to the lower energy states. Imagine that our
Fermions form $m := N / n$ teams, where each team has $n$ members with distinct denominations (or ranks, or positions)
and there is no mutual interaction (repulsion) whatsoever within each team; so the only interactions are the inter-team
interactions. Furthermore, the pattern of interactions between any two teams is independent of the choice of teams,
{\it i.e.}, the teams are essentially indistinguishable. Given a pair of teams, we stipulate at most two possible
types of inter-team interactions: interaction of potency $b$ between the members of similar denominations and interaction
of potency $a$ between the members of dissimilar denominations (which exist only when $n \geq 2$). So the $n$ denominations
are also essentially indistinguishable. If the Fermions in such a configuration are regrouped by denominations, then we obtain
$n$ groups containing $m$ Fermions each, where for each group, the intra-group interactions are of the same potency $b$ and
the total interaction potency between two groups is (the even integer) $m (m - 1) a$.
The graphic in Fig. 5 below illustrates the interactions between two teams when $n = 3$, $b = 2$ and $a  = 1$.

\begin{figure}[h!]
\centering
\includegraphics[scale=0.4]{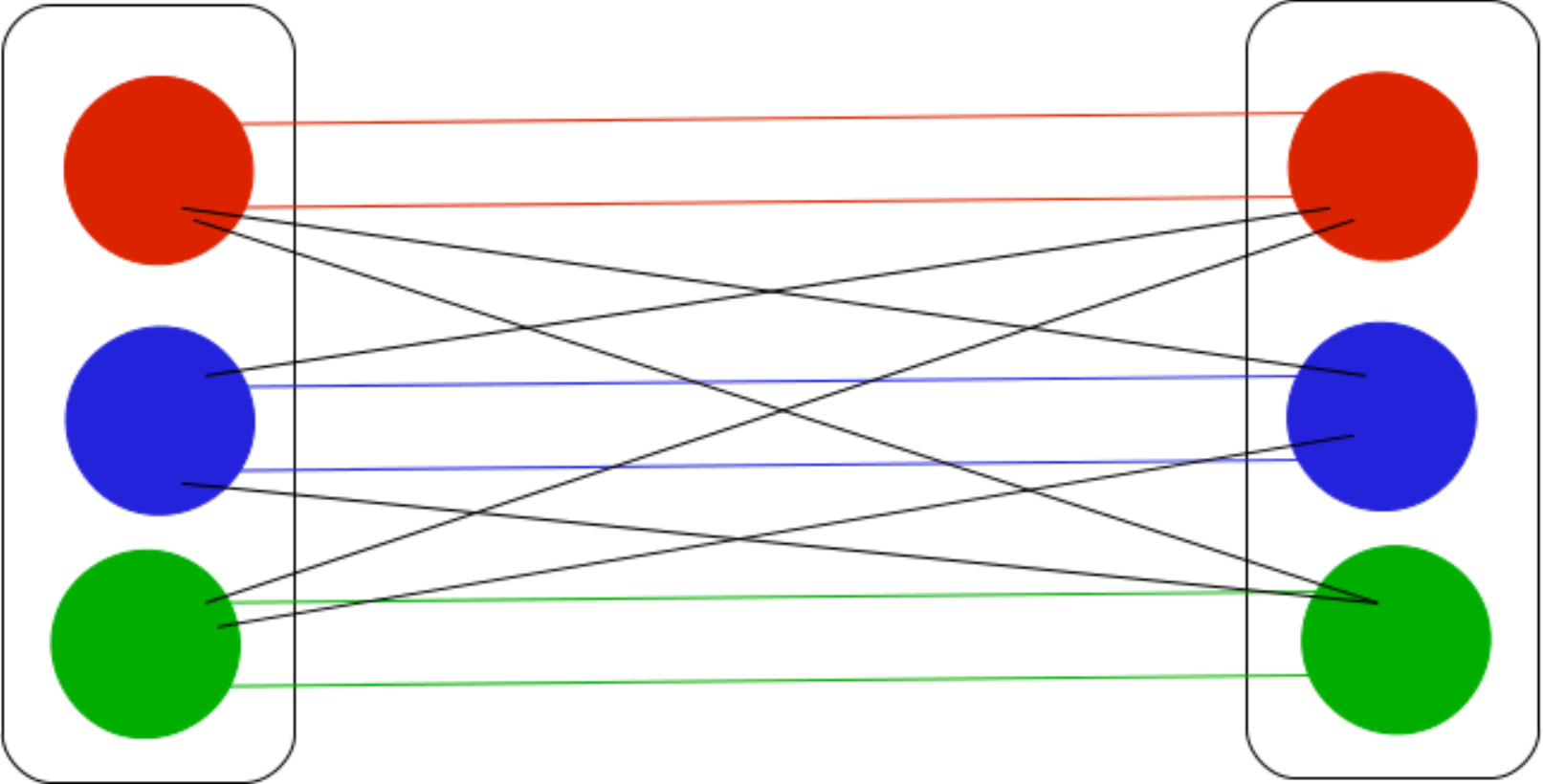}
\caption{\small Interactions between two teams when $(n , b, a) = (3, 2, 1)$.}
\end{figure}

We will refer to members of the pool of possible configurations described above as {\it balanced}. Note that for $n = 1$,
{\it i.e.}, $\nu = 1 / (2 p + 1)$, there is just one balanced configuration: namely, the familiar Laughlin configuration.  Further, we define {\it minimal configurations} to be those balanced configurations where max$\{a, b \}$ is at its least possible value for a given filling factor. One can show for example that the two diagrams in Fig. 3  above correspond to the only two minimal configurations involving $6$ Fermions in the IQL state with filling factor ${^2/_5}$. Of course, for $n = 1$ and arbitrary $p$, the standard Laughlin configuration is indeed a minimal configuration according to our definition.   Moreover, since the total number of correlations between any two teams of $n$ Fermions in a minimal configuration is the even integer $(2 p - 1) n^2 \pm n$, one may regard each team as a `super particle', in which case the interaction configuration of these $m$ super particles is of Laughlin type.

We conclude with the following result on the existence of minimal configurations.

\noindent \underline {\bf Theorem 4}:  Minimal configurations always exist, \emph{i.e.}, have a nonzero symmetric correlation factor $G$, for all filling factors $\nu < {^1/_2}$ of the form $\frac{n}{2pn\pm1}<\frac{1}{2}$ where $N$ is a multiple of $n$.\\
\noindent \underline{\bf Proof}: First note that $\frac{d}{m-1}=\frac{d}{N/n-1}=(n-1)a+b$ for balanced configurations, where $d$ denotes the number of interactions of each particle. Thus the condition $L = 0$ translates in this case to the requirement
\[(n - 1) a + b = (2 p - 1) n \pm 1. \;\;\;\;\;\;\;\;\;\;\;\;\;(*) \]
As long as either
$n = 1$ or $a$ and $b$ both are positive, Theorem 3 above, together with its corollary, guarantees an existent
configuration. If $a$, $b$ both are even integers, then Theorem 2 implies that we have an existent configuration in this case.
(On the other hand, if $a = 0$ and $b$ is odd and thus $n$ is even, then it can be shown that in fact there is a non-existent configuration, \emph{i.e.}, one in which $G=0$.)
Now observe that the solutions to our constraint $(*)$ such that $\max\{a,b\}$ is minimal and $n\geq 2$ are given by
\[ (a, \, b) \, = \, \left \{ \begin{array}{ll} (2p,2p-n+1),(2 p - 1, \, 2 p) & \mbox{if $\nu = n / (2 p n + 1)$ and} \\
(2 p - 1, \, 2 p - 2) & \mbox{if $\nu = n / (2 p n - 1)$,} \end{array} \right . \]
where if $n = 2$, then there is also the `minimal' solution $(a,b)=(2p-2,2p-1)$ in the second case (note that the first solution in the first case above does not exist if $n>2p+1$). Since $a$, $b$ both are clearly positive (or even) in each of the preceding cases, this implies the existence of minimal configurations. $\Box$ \\

\bibliographystyle{plain}
\bibliography{mqsbiblioIQL_abbrv}

\providecommand{\noopsort}[1]{}\providecommand{\singleletter}[1]{#1}%
\begin{thebibliography}{10}

\bibitem{AS}
P.~Alexandersson and B.~Shapiro.
\newblock Discriminants, symmetrized graph monomials, and sums of squares.
\newblock {\em Experiment. Math.}, 21:353--361, 2012.

\bibitem{BQ}
A.~T. Benjamin, Jennifer~J. Quinn, John~J. Quinn, and A.~W\'{o}js.
\newblock Composite {F}ermions and integer partitions.
\newblock {\em J. Combin. Theory Ser. A}, 95:390--397, 2001.

\bibitem{BDP}
A.~E. Brouwer, J.~Draisma, and M.~Popoviciu.
\newblock The degrees of a system of parameters of the ring of invariants of a
  binary form.
\newblock http://arxiv.org/abs/1404.5722.

\bibitem{D}
J.~Dixmier.
\newblock Quelques r$\acute{e}$sultats et conjectures concernant les
  s$\acute{e}$ries de {P}oincar$\acute{e}$ des invariants des formes binaires.
\newblock In {\em S$\acute{e}$minaire d'alg$\grave{e}$bre Paul Dubreil et
  Marie-Paule Malliavin, 36$\grave{e}$me ann$\acute{e}$e (Paris, 1983-1984)},
  pages 127--160. Springer, Berlin, 1985.
\newblock Lecture Notes in Math., 1146.

\bibitem{E}
E.~B. Elliot.
\newblock {\em An Introduction to the Algebra of Quantics}.
\newblock Chelsea Publishing Company, New York, 1964.
\newblock Second edition (1913), reprint.

\bibitem{GY}
J.~H. Grace and A.~Young.
\newblock {\em The Algebra of Invariants}.
\newblock Chelsea Publishing Company, New York, 1964.
\newblock (1903), reprint.

\bibitem{GM}
C.~Greenhill and B.~D. McKay.
\newblock Asymptotic enumeration of sparse multigraphs with given degrees.
\newblock {\em SIAM J. Discrete Math.}, 27:2064--2089, 2013.

\bibitem{Jain89}
J.~K. Jain.
\newblock Composite-{F}ermion approach for the fractional quantum {H}all
  effect.
\newblock {\em Phys. Rev. Lett.}, 63:199--202, Jul 1989.

\bibitem{Jain90}
J.~K. Jain.
\newblock Theory of the fractional quantum {H}all effect.
\newblock {\em Phys. Rev. B}, 41:7653--7665, Apr 1990.

\bibitem{KR}
J.~P.~S. Kung and G.-C. Rota.
\newblock The invariant theory of binary forms.
\newblock {\em Bull. Amer. Math. Soc.}, 10:27--85, Jan 1984.

\bibitem{Laughlin83}
R.~B. Laughlin.
\newblock Anomalous quantum {H}all effect: An incompressible quantum fluid with
  fractionally charged excitations.
\newblock {\em Phys. Rev. Lett.}, 50:1395--1398, May 1983.

\bibitem{Moore91}
G.~Moore and N.~Read.
\newblock Nonabelions in the fractional quantum {H}all effect.
\newblock {\em Nucl. Phys. B}, 360(2–3):362--396, 1991.

\bibitem{Quinn14}
J.~J. Quinn.
\newblock Constructing trial wave functions for a many electron system confined
  to a quantum well in a strong magnetic field.
\newblock {\em Waves Random Complex Media}, 24(3):279--285, 2014.

\bibitem{Quinn09}
J.~J. Quinn, A.~W{\'o}js, K.-S. Yi, and G.~Simion.
\newblock The hierarchy of incompressible fractional quantum {H}all states.
\newblock {\em Phys. Rep.}, 481(3–4):29--81, 2009.

\bibitem{QY}
J.~J. Quinn and K.-S. Yi.
\newblock {\em Solid State Physics: Principles and Modern Applications}.
\newblock Springer-Verlag, Berlin, 2009.

\bibitem{SZ}
O.~Zariski and P.~Samuel.
\newblock {\em Commutative Algebra}, volume I and II.
\newblock Springer, New York, 1976.
\newblock Graduate Texts in Mathematics.

\end{thebibliography}
\end{document}